\documentclass[iop]{emulateapj} 

\def \lp{\>\> .}  \def \lc{\>\> ,} \def \arcsec{\hbox{$^{\prime\prime}$}}
\def \arcmin{\hbox{$^{\prime}$}}

\def \nh3{NH$_3$}
\def \n2h{N$_2$H$^+$}

\def \nh2{n_{H_2}}
\def \nh1{n_{HI}}

\def \h2{H$_2$}

\def \c2{cm$^{-2}$}

\def \be{\begin{equation}}
\def \ee{\end{equation}}
\def \ef{\end{figure}}

\usepackage{amsmath}

\slugcomment{To appear in the Astrophysical Journal}

\shorttitle{The relation between gas and dust in the Taurus Molecular Cloud} \shortauthors{Pineda, Goldsmith, Chapman, Snell, Li, Cambr\'esy \& Brunt}

\begin{document}


\title{The relation between gas and dust in the Taurus Molecular Cloud}

\author{Jorge L. Pineda$^{1}$, Paul F. Goldsmith$^{1}$, Nicholas Chapman$^{1}$, Ronald L. Snell$^{2}$, Di Li$^{1}$, Laurent Cambr\'esy$^{3}$, and Chris Brunt$^{4}$}  
\affil{$^{1}$Jet Propulsion Laboratory, California Institute of Technology, 4800 Oak Grove Drive, Pasadena, CA 91109-8099, USA\\
$^{2}$Department of Astronomy, LGRT 619, University of Massachusetts, 710 North Pleasant Street, Amherst, MA 01003, USA\\
$^{3}$Observatoire Astronomique de Strasbourg, 67000 Strasbourg, France\\
$^{4}$Astrophysics Group, School of Physics, University of Exeter, Stocker Road, Exeter, EX4 4QL, UK}

\email{Jorge.Pineda@jpl.nasa.gov}

\begin{abstract}

We report a study of the relation between dust and gas over a 100
deg$^{2}$ area in the Taurus molecular cloud.  We compare the H$_2$
column density derived from dust extinction with the CO column density
derived from the $^{12}$CO and $^{13}$CO $J= 1 \to 0$ lines.  We
derive the visual extinction from reddening determined from 2MASS
data.  The comparison is done at an angular size of 200\arcsec,
corresponding to 0.14\,pc at a distance of 140\,pc.  We find that the
relation between visual extinction $A_{\rm V}$ and $N({\rm CO})$ is
linear between $A_{\rm V} \simeq 3$ and 10\,mag in the region
associated with the B213--L1495 filament.  In other regions the linear
relation is flattened for $A_{\rm V} \gtrsim$ 4\,mag.  We find that
the presence of temperature gradients in the molecular gas affects the
determination of $N({\rm CO})$ by $\sim$30--70\% with the largest
difference occurring at large column densities.  Adding a correction
for this effect and accounting for the observed relation between the
column density of CO and CO$_2$ ices and $A_{\rm V}$, we find a linear
relationship between the column of carbon monoxide and dust for
observed visual extinctions up to the maximum value in our data
$\simeq 23$\, mag.  We have used these data to study a sample of dense
cores in Taurus. Fitting an analytical column density profile to these
cores we derive an average volume density of about
$1.4\times10^{4}$\,cm$^{-3}$ and a CO depletion age of about
$4.2\times10^{5}$\,years.  At visual extinctions smaller than
$\sim$3\,mag, we find that the CO fractional abundance is reduced by
up to two orders of magnitude. The data show a large scatter
suggesting a range of physical conditions of the gas.  We estimate the
H$_2$ mass of Taurus to be about $1.5\times10^{4}$\,M$_{\odot}$,
independently derived from the $A_{\rm V}$ and $N({\rm CO})$ maps. We
derive a CO integrated intensity to H$_2$ conversion factor of about
2.1$\times10^{20}$\,cm$^{-2}$(K km s$^{-1}$)$^{-1}$, which applies
even in the region where the [CO]/[H$_2$] ratio is reduced by up to
two orders of magnitude.  The distribution of column densities in our
Taurus maps resembles a log--normal function but shows tails at large
and low column densities. The length scale at which the high--column
density tail starts to be noticeable is about 0.4\,pc.

\end{abstract}

\keywords{ISM: molecules --- ISM: structure}

\section{Introduction}
\label{sec:introduction}

Interstellar dust and gas provide the primary tools for tracing the
structure and determining the mass of extended clouds as well as more
compact, dense regions within which new stars form.  The most
fundamental measure of the amount material in molecular clouds is the
number of H$_2$ molecules along the line of sight averaged over an
area defined by the resolution of the observations, the H$_2$ column
density, $N({\rm H}_2)$. Unfortunately, H$_2$ has no transitions that
can be excited under the typical conditions of molecular clouds, and
therefore it cannot be directly observed in such regions.  We have to
rely on indirect methods to determine $N({\rm H}_2)$. Two of the most
common methods are observations of CO emission and dust extinction.

Carbon monoxide (CO) is the second most abundant molecular species
(after H$_2$) in the Universe. Observations of $^{12}$CO and $^{13}$CO
together with the assumption of local thermodynamic equilibrium (LTE)
and moderate $^{13}$CO optical depths allow us to determine $N({\rm
CO})$ and, assuming an [CO]/[H$_2$] abundance ratio, we can obtain
$N({\rm H}_2)$.  This method is, however, limited by the sensitivity
of the $^{13}$CO observations and therefore is only able to trace
large column densities. \citet{Goldsmith2008} used a 100 square degree
map of $^{12}$CO and $^{13}$CO in the Taurus molecular cloud to derive
the distribution of $N({\rm CO})$ and $N({\rm H}_2)$.  By binning the
CO data by excitation temperature, they were able to estimate the CO
column densities in individual pixels where $^{12}$CO but not
$^{13}$CO was detected.  The pixels where neither $^{12}$CO or
$^{13}$CO were detected were binned together to estimate the average
column density in this portion of the cloud.

Extensive work has been done to assess the reliability of CO as a
tracer of the column of H$_2$ molecules \citep[e.g.][]{FLW1982,
Langer1989}.  It has been found that $N({\rm CO})$ is not linearly
correlated with $N({\rm H}_2)$, as the former quantity is sensitive to
chemical effects such as CO depletion at high volume densities
\citep[][]{Kramer1999,Caselli1999,Tafalla2002} and the competition
between CO formation and destruction at  low-column densities
\citep[e.g.][]{vanDishBlack88,Visser2009}. Moreover, temperature
gradients are likely present in molecular clouds
\citep[e.g.][]{Evans2001} affecting the correction of $N({\rm CO})$
for optical depth effects.


The H$_2$ column density can be independently inferred by measuring
the optical or near--infrared light from background stars that has
been extincted by the dust present in the molecular cloud
\citep{Lada1994,Cambresy1999,Dobashi2005}.  This method is often
regarded as one of the most reliable because it does not depend
strongly on the physical conditions of the dust.  But this method is
not without some uncertainty.  Variations in the total to selective
extinction and dust--to--gas ratio, particularly in denser clouds like
those in Taurus, may introduce some uncertainty in the conversion of
the infrared extinction to gas column density \citep{Whittet2001}.
Dust emission has been also used to derive the column density of H$_2$
\citep{Langer1989}. It is, however, strongly dependent on the dust
temperature along the line of sight, which is not always well
characterized and difficult to determine.  Neither method provides
information about the kinematics of the gas.

It is therefore of interest to compare column density maps derived
from $^{12}$CO and $^{13}$CO observations with dust extinction
maps. This will allow us to characterize the impact of
chemistry and saturation effects in the derivation of $N({\rm CO})$ and
$N({\rm H}_2)$ while testing theoretical predictions of the physical
processes that cause these effects.

As mentioned before, CO is frozen onto dust grains in regions of
relatively low temperature and larger volume densities
\citep[e.g.][]{Kramer1999,Tafalla2002,Bergin2002}.  In dense cores,
the column densities of C$^{17}$O \citep{Bergin2002} and C$^{18}$O
\citep{Kramer1999,Alves1999,Kainulainen2006} are observed to be
linearly correlated with $A_{\rm V}$ up to $\sim$10\,mag.  For larger
visual extinctions this relation is flattened with the column density
of these species being lower than that expected for a constant
abundance relative to H$_2$.  These authors showed that the C$^{17}$O
and C$^{18}$O emission is optically thin even at visual extinctions
larger than 10\,mag and therefore the flattening of the relation
between their column density and $A_{\rm V}$ is not due to optical
depths effects but to depletion of CO onto dust grains.  These
observations suggest drops in the relative abundance of C$^{18}$O {\it
averaged along the line-of-sight} of up to a factor of $\sim$3 for
visual extinctions between 10 and 30\,mag. A similar result has been
obtained from direct determinations of the column density of CO--ices
based on absorption studies toward embedded and field stars
\citep{Chiar1995}. At the center of dense cores, the [CO]/][H$_2$]
ratio is expected to be reduced by up to five orders of magnitude
\citep{Bergin1997}. This has been confirmed by the comparison between
observations and radiative transfer calculations of dust continuum and
C$^{18}$O emission in a sample of cores in Taurus
\citep{Caselli1999,Tafalla2002}.  The amount of depletion is not only
dependent on the temperature and density of the gas, but is also
dependent on the timescale.  Thus, determining the amount of depletion
in a large sample of cores distributed in a large area is important
because it allow us to determine the chemical age of the entire Taurus
molecular cloud while establishing the existence of any systematic
spatial variation that can be a result of a large--scale dynamical
process that lead to its formation.

At low column densities ($A_{\rm V}\lesssim$ 3\,mag, $N({\rm
CO})\lesssim$\,10$^{17}$\,cm$^{-2}$) the relative abundance of CO and
its isotopes are affected by the relative rates of formation and
destruction, carbon isotope exchange and isotope selective
photodissociation by far-ultraviolet (FUV) photons.  These effects can
reduce [CO]/[H$_2$] by up to three orders of magnitude
\citep[e.g.][]{vanDishBlack88,Liszt2007,Visser2009}.  This column
density regime has been studied in dozens of lines-of-sight using UV
and optical absorption
\citep[e.g.][]{Federman1980,Sheffer2002,Sonnentrucker2003,Burgh2007}
as well as in absorption toward mm-wave continuum sources
\citep{Liszt1998}.  The statistical method presented by
\citet{Goldsmith2008} allows the determination of CO column densities
in several hundred thousand positions in the periphery of the Taurus
molecular cloud with $N({\rm CO}) \simeq 10^{14}-10^{17}$\,cm$^{-3}$.
A comparison with the visual extinction will provide a coherent
picture of the relation between $N({\rm CO})$ and $N({\rm H}_2)$ from
diffuse to dense gas in Taurus. These results can be compared with
theoretical predictions that provide constraints in physical
parameters such as the strength of the FUV radiation field, etc.

Accounting for the various mechanisms affecting the [CO]/[H$_2$]
relative abundance allows the determination of the H$_2$ column density
that can be compared with that derived from $A_{\rm V}$ in the Taurus
molecular cloud.  It has also been suggested that the total molecular
mass can be determined using only the integrated intensity of the
$^{12}$CO $J = 1 \to 0$ line together with the empirically--derived
CO-to-H$_2$ conversion factor ($X_{\rm CO} \equiv N({\rm H}_2)/I_{\rm
CO})$.  The $X_{\rm CO}$ factor is thought to be dependent on the
physical conditions of the CO--emitting gas \citep{MaloneyBlack88} but
it has been found to attain the canonical value for our Galaxy even in
diffuse regions where the [CO]/][H$_2$] ratio is strongly
affected by CO formation/destruction processes \citep{Liszt2007}.  The
large--scale maps of $N({\rm CO})$ and $A_{\rm V}$ also allow us to
assess whether there is H$_2$ gas that is not traced by CO. This
so-called ``dark gas'' is suggested to account for a substantial
fraction of the total molecular gas in our Galaxy \citep{Grenier2005}.


 Numerical simulations have shown that the probability density
function (PDF) of volume densities in molecular clouds can be fitted
by a log-normal distribution
\citep[e.g.][]{Ostriker2001,Nordlund1999,Li2004,Klessen2000}.  The
shape of the distribution is expected to be log-normal as
multiplicative effects determine the volume density of a molecular
cloud \citep{Passot1998,Vazquez-Semadeni2001}.  A log-normal function
can also describe the distribution of column densities in a molecular
cloud \citep{Ostriker2001,Vazquez-Semadeni2001}. For some molecular
clouds the column density distribution can be well fitted by a
log--normal \citep[e.g.][]{Wong2008,Goodman2009}.
A study by \citet{Kainulainen2009}, however, showed that in a larger
sample of molecular complexes the column density distribution shows
tails at low and large column densities.
The presence of tails at large column
densities seems to be linked to active star--formation in clouds.  The
$A_{\rm V}$ and CO maps can be used to determine the distribution of
column densities at large scales while allowing us to study variations
in its shape in regions with different star--formation activity within
Taurus.

In this paper, we compare the CO column density derived using the
$^{12}$CO and $^{13}$CO data from \citealt{Narayanan2008} (see also
\citealt{Goldsmith2008}) with a dust extinction map of the Taurus
molecular cloud.  The paper is organized as follows: In
Section~\ref{sec:nh_2-map-derived} we describe the derivation of the
CO column density in pixels where both $^{12}$CO and $^{13}$CO were
detected, where $^{12}$CO but not $^{13}$CO was detected, as well as
in the region where no line was detected in each individual pixel.  In
Section~\ref{sec:comp-betw-a_v} we make pixel--by--pixel comparisons
between the derived $N({\rm CO})$ and the visual extinction for the
large and low column density regimes.  In
Section~\ref{sec:mass-taur-molec} we compare the total mass of Taurus
derived from $N({\rm CO})$ and $A_{\rm V}$. We also study how good the
$^{12}$CO luminosity together with a CO-to-H$_2$ conversion factor can
determine the total mass of a molecular cloud. We study the
distribution of column densities in Taurus in
Section~\ref{sec:column-prob-dens}. We present a summary of our
results in Section~\ref{sec:conclusions}.

\section{The N(H$_2$) map derived from $^{12}$CO and $^{13}$CO}
\label{sec:nh_2-map-derived}

\begin{figure}[t]
\includegraphics[width=0.48\textwidth,angle=0]{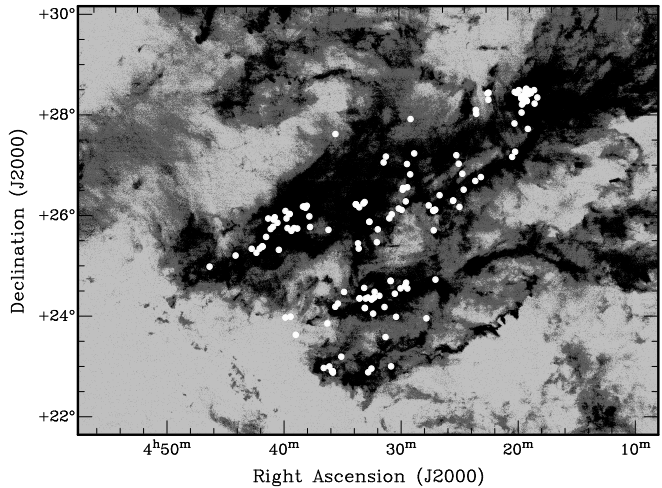}
     \caption{Mask regions defined in the Taurus Molecular Cloud.
     Mask 2 is shown in black, Mask 1 in dark gray, and Mask 0 in
     light gray. We also show the 156  stellar members of
     Taurus compiled by \citet{Luhman2006} as white circles.
     }\label{fig:masks}
\end{figure}

In the following we derive the column density of CO using the FCRAO
14--m $^{12}$CO and $^{13}$CO observations presented by
\citealt{Narayanan2008} (see also \citealt{Goldsmith2008}).  In this
paper we use data corrected for error beam pick-up using the method
presented by \citet{Bensch2001}. The correction procedure is described
in Appendix~\ref{sec:error-beam-corr}.  The correction for error beam
pick--up improves the calibration by 25--30\%.  We also improved the
determination of $N({\rm CO})$ compared to that presented by
\citet{Goldsmith2008} by including an updated value of the spontaneous
decay rate and using an exact numerical rather than approximate
analytical calculation of the partition function.  The values of the
CO column density are about $\sim$20\% larger than those presented by
\citet{Goldsmith2008}.  Following \citet{Goldsmith2008}, we define
Mask\,2 as pixels where both $^{12}$CO and $^{13}$CO are detected,
Mask\,1 as pixels where $^{12}$CO is detected but $^{13}$CO is not,
and Mask\,0 as pixels where neither $^{12}$CO nor $^{13}$CO are
detected.  We consider a line to be detected in a pixel when its
intensity, integrated over the velocity range between
0--12\,km\,s$^{-1}$, is at least 3.5 times larger than the rms noise
over the same velocity interval. We show the mask regions in
Figure~\ref{fig:masks}.  The map mean rms noise over this velocity
range is $\sigma_{T^*_{\rm int}}$=0.53\,K\,km\,s$^{-1}$ for $^{12}$CO
and $\sigma_{T^*_{\rm int}}$=0.23\,K\,km\,s$^{-1}$ for $^{13}$CO. The
map mean signal-to-noise ratio is 9 for $^{12}$CO and 7.5 for
$^{13}$CO.  Note that these values differ slightly from those
presented by \citet{Goldsmith2008}, as the correction for error beam
pick--up produces small changes in the noise properties of the data.

\subsection{CO Column Density in Mask 2}
\label{sec:co-column-density_mask2}

\subsubsection{The Antenna Temperature}

When we observe a given direction in the sky, the antenna temperature
we measure is proportional to the convolution of the brightness of the
sky with the normalized power pattern of the antenna.  Deconvolving
the measured set of antenna temperatures is relatively difficult,
computationally expensive, and in consequence rarely done.  The
simplest approximation that is made is that the observed antenna
temperature is that coming from a source of some arbitrary size,
generally that of the main beam, or else a larger region.  It is
assumed that the measured antenna temperature can be corrected for the
complex antenna response pattern and its coupling to the (potentially
nonuniform) source by an efficiency, characterizing the coupling to
the source.  This is often taken to be $\eta_{\rm mb}$, the coupling
to an uniform source of size which just fills the main lobe of the
antenna pattern. This was the approach used by
\citet{Goldsmith2008}. In Appendix~\ref{sec:error-beam-corr} we
discuss an improved technique which corrects for the error pattern of
the telescope in the Fourier space. This technique introduces a
``corrected main--beam temperature scale'', $T_{\rm mb,c}$.  
We can write  the main--beam corrected temperature as


\begin{equation}   T_{\rm mb,c} = T_0 \left [\frac{1}{e^{T_0/T_{\rm ex}} - 1} -
\frac{1}{e^{T_0/T_{bg}} - 1} \right ] \left (1 - e^{-\tau} \right),
\label{eq:14}
\end{equation}
 where $T_0$ = $h\nu/k$, $T_{\rm ex}$ is the excitation temperature of
the transition, $T_{bg}$ is the background radiation temperature, and
$\tau$ is the optical depth.  This equation applies to a given
frequency of the spectral line, or equivalently, to a given velocity,
and the optical depth is that appropriate for the frequency or
velocity observed.


%

If we assume that the excitation temperature is independent of
velocity (which is equivalent to an assumption about the uniformity of
the excitation along the line--of--sight) and integrate over velocity we
obtain 
\begin{equation}
\int  T_{\rm mb,c}(v) dv = \frac{T_0 C(T_{\rm ex})}{e^{T_0/T_{\rm ex}} -
1}\int(1 - e^{-\tau(v)})dv\lc \label{eq:15}
\end{equation} 
where we have included explicitly the dependence of the corrected
main--beam temperature and the optical depth on velocity. The function
$C(T_{\rm ex})$, which is equal to unity in the limit $T_{bg} \to 0$,
is given by

\begin{equation}
\label{eq:8}
C(T_{\rm ex})=\left (1-\frac{e^{T_0/T_{\rm ex}} - 1}{e^{T_0/T_{bg}} - 1} \right).
\end{equation}
\begin{figure}
\includegraphics[scale=0.43]{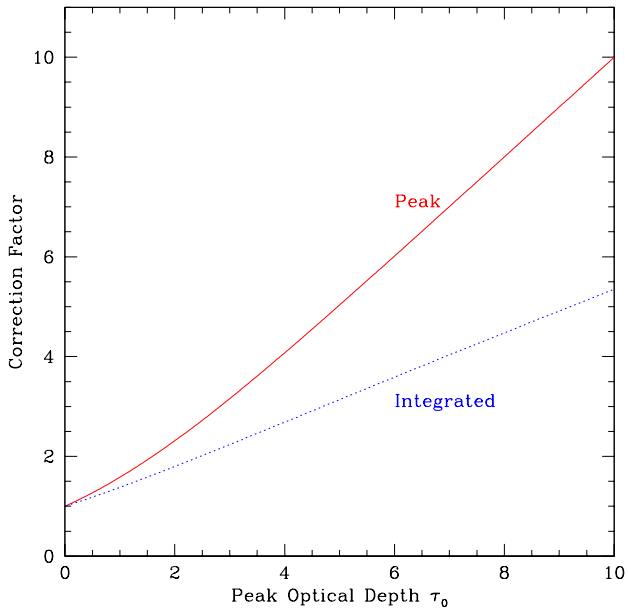}
\caption{Correction factors for the relation between integrated
main--beam temperature and upper level column density for a Gaussian
 velocity distribution of the optical depth.  The dotted (blue)
curve shows the correction factor obtained using integrals of
functions of the optical depth as given by Equation\,\ref{cf1}.  The
solid (red) curve shows the correction factor employing the peak
values of the functions, given by Equation\,\ref{cf2}.
}\label{taucorrcomp}
\end{figure}

\subsubsection{The Optical Depth}

The optical depth is determined by the difference in the populations
of the upper and lower levels of the transition observed.  If we
assume that the line--of--sight is characterized by upper and lower
level column densities, $N_U$ and $N_L$, respectively, the optical
depth is given by

\begin{equation} \tau = \frac{h\nu_0}{c} \phi(\nu)\left [N_LB_{LU} - N_UB_{UL} \right],
\end{equation} 
where $\nu_0$ is the frequency of the transition, $\phi(\nu)$ is the
line profile function, and the $B$'s are the Einstein B-coefficients.
The line profile function is a function of the frequency and describes
the relative number of molecules at each frequency (determined by
relative Doppler velocity).  It is normalized such that
$\int\phi(\nu)d\nu$ = 1.  For a Gaussian line profile, the line
profile function at line center is given approximately by $\phi(\nu_0)
= 1/\delta\nu_{\rm FWHM}$, where $\delta\nu_{\rm FWHM}$ is the full
width at the half maximum of the line profile.


 We have assumed that the excitation temperature is uniform along
	the line of sight.  Thus, we can define the excitation
	temperature in terms of the upper and lower level column
	densities, and we can write 

\begin{equation}
\frac{N_U}{N_L} = \frac{g_U}{g_L}e^{-T_0/T_{\rm ex}},
\end{equation}
where the $g$'s are the statistical weights of the two levels.
The relationship between the $B$ coefficients,
\begin{equation}
g_U B_{UL} = g_L B_{LU},
\end{equation}
lets us write
\begin{equation}
\tau(\nu) = \frac{h \nu_o B_{UL}\phi(\nu) N_U}{c} \left [e^{T_0/T_{\rm ex}} - 1\right  ].
\end{equation}
Substituting the relationship between the $A$ and $B$ coefficients, 
\begin{equation}
A_{UL} = B_{UL}\frac{8 \pi h\nu_0^3}{c^3}\lc
\end{equation}
gives us
\begin{equation}
\tau(\nu) = \frac{c^2 A_{UL} \phi(\nu) N_U}{8 \pi {\nu_0}^2}\left [e^{T_0/T_{\rm ex}} - 1 \right  ].
\end{equation}
If we integrate both sides of this equation over a range of frequencies encompassing the entire spectral line of interest, we find
\begin{equation}
\int \tau(\nu)d\nu = \frac{c^2 A_{UL} N_U}{8 \pi {\nu_0}^2} \left [e^{T_0/T_{\rm ex}} - 1 \right  ].
\end{equation} 
\begin{figure}
\centering
\includegraphics[width=0.47\textwidth]{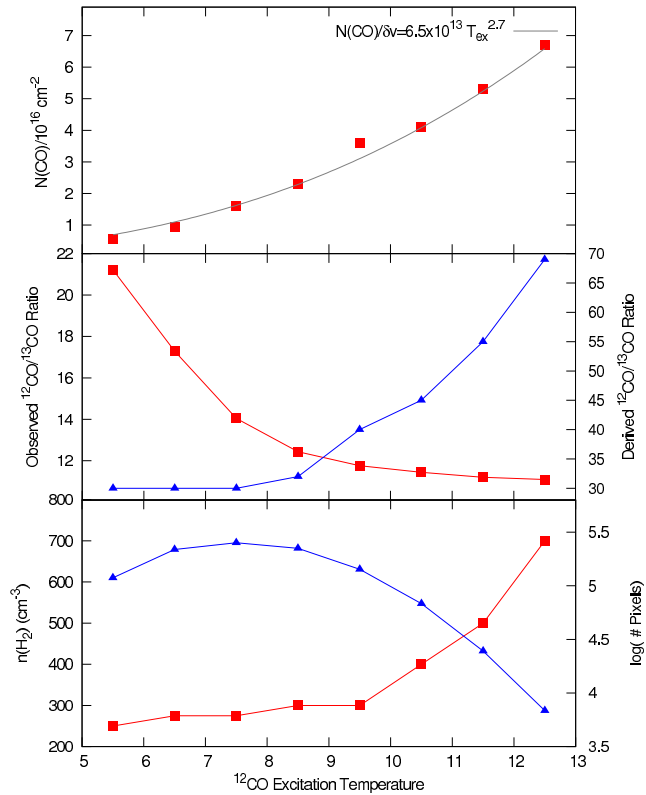}
\caption{Parameters of Mask 1 binned by $^{12}$CO excitation
temperature $T_{\rm ex}$. The bottom panel shows the derived H$_2$
density ({\it left-hand scale, squares}) and the number of pixels in
each $T_{\rm ex}$ bin ({\it right-hand scale, triangles}). The most
common $T_{\rm ex}$ values are between $5-9$\,K. The middle panel shows the
observed ({\it left-hand scale, squares}) and derived ({\it right-hand
scale, triangles}) $^{12}$CO/$^{13}$CO ratio. The top panel shows the
derived $^{12}$CO column density assuming a line width of
1\,Km\,s$^{-1}$. The H$_2$ density, $^{12}$CO column density and derived
$^{12}$CO/$^{13}$CO ratio increase monotonically as a function of
$^{12}$CO excitation temperature.}\label{fig:ron_results}
\end{figure}

\begin{deluxetable*}{lcrccc}
\tabletypesize{\scriptsize} \centering \tablecolumns{15} \small
\tablewidth{0pt}
\tablecaption{$^{12}$CO Excitation Temperature Bins in Mask 1 and Best Estimates of Their Characteristics}
\tablenum{1}
\tablehead{\colhead{$T_{\rm ex}$} & \colhead{$^{12}$CO/$^{13}$CO}  & \colhead{Number} & \colhead{$n({\rm H}_2)$} & \colhead{$N({\rm CO})/\delta v$} &  \colhead{$^{12}$CO/$^{13}$CO}\\
\colhead{(K)} & \colhead{Observed}  & \colhead{of Pixels} & \colhead{(cm$^{-3}$)} & \colhead{($10^{16}$ cm$^{-2}$/km s$^{-1}$)} &  \colhead{Abundance Ratio}}
\startdata
  5.5....................... & 21.21 &118567  & 250  & 0.56 & 30  \\
  6.5....................... & 17.29 &218220 & 275 &0.95 &  30 \\
  7.5....................... & 14.04 &252399 & 275 &1.6  &  30 \\
  8.5....................... & 12.43 &223632 & 300 &2.3  &  32 \\
  9.5....................... & 11.76 &142525 & 300 &3.6  &  40 \\
  10.5......................  & 11.44 &68091  &400 &4.1  &  45 \\
  11.5......................  & 11.20 &24608  &500 &5.3  &  55 \\
  12.5......................  & 11.09 &6852   &700 &6.7  &  69
\enddata
\end{deluxetable*}

\subsubsection{Upper Level Column Density}

It is generally more convenient to describe the optical depth in terms of the velocity offset relative to that of the nominal line center.  The incremental frequency and velocity are related through
$dv = (c/\nu_0)d\nu$, and hence $\int \tau(\nu)d\nu = (c/\nu_0)\int \tau(v)dv$.  Thus we obtain
\begin{equation}\label{eq:2}
\int \tau(v)dv = \frac {c^3 A_{UL} N_U} {8 \pi {\nu_0}^3}\left [e^{T_0/T_{\rm ex}} - 1 \right].
\end{equation} 
We can rewrite this as
\begin{equation}\label{eq:16}
\frac{1}{e^{T_0/T_{\rm ex}} - 1} = \frac{c^3 A_{UL} N_U}{8 \pi {\nu_0}^3} \frac{1}{\int \tau(v)dv}.
\end{equation} 
Substituting this into Equation~(\ref{eq:15}), we can write an
expression for the upper level column density as
\begin{equation}\label{eq:17}
N_U = \frac{8 \pi k \nu_0^2}{ h c^3 A_{UL} C(T_{\rm ex})} \left [
\frac{\int \tau(v)dv} {\int(1 - e^{-\tau(v)})dv} \right]\int  T_{\rm mb,c}(v) dv.
\end{equation} 
For the calculation of the $^{13}$CO column densities
(Section~\ref{sec:total-13co-column}) we use a value for the Einstein
A-coefficient of $A_{UL}=$6.33$\times10^{-8}$ s$^{-1}$
\citep{Goorvitch1994}.

In the limit of optically thin emission for which $\tau(v)$ $\ll$ 1
for all $v$, and neglecting the background term in
Equation~(\ref{eq:8})\footnote{This usually does not result in a
significant error since in LTE even in dark clouds $T_{\rm ex}$ is
close to 10 K as compared to $T_{bg}$ = 2.7 K.  Since $T_{bg}$ is
significantly less than $T_0$, the background term is far from the
Rayleigh--Jeans limit further reducing its magnitude relative to that
of the first term. }, the expression in square brackets is unity and
we regain the much simpler expression

\begin{equation}\label{eq:1}
N_U(thin) = \frac{8 \pi k \nu_0^2}{h c^3 A_{UL}}\int T_{\rm mb,c}(v)
dv \lp \end{equation}  We will, however, use the general form of
$N_U$ given in Equation~(\ref{eq:17}) for the determination of the CO
column density.

We note that the factor in square brackets in Equation~(\ref{eq:17})
involves the {\it integrals} of functions of the optical depth over
velocity, not just the functions themselves.  There is a difference,
which is shown in Figure~\ref{taucorrcomp}, where we plot the two
functions
\begin{equation}
\label{cf1}
CF(integral) = \frac{\int \tau(v)dv}{\int(1 - e^{-\tau(v)})dv}\lc
\end{equation}
and
\begin{equation}
\label{cf2}
CF(peak) = \frac{\tau_0}{1 - e^{-\tau_0}}\lc \end{equation} as a function of the
peak optical depth $\tau_0$.  There is a substantial difference at
high optical depth, which reflects the fact that the line center has
the highest optical depth so that using this value rather than the
integral tends to overestimate the correction factor.

\subsubsection{Total $^{13}$CO column densities derived from $^{13}$CO  and $^{12}$CO observations.  }
\label{sec:total-13co-column}

In LTE, the column density of the upper level ($J=1$) is related to the total
$^{13}$CO column density by

\begin{equation}
\label{eq:9}
N_{\rm ^{13}CO}=N_U \frac{Z}{(2J+1)} e^\frac{h B_0 J(J+1)}{K T_{\rm ex}}
\end{equation}
where $B_0$ is the rotational constant of $^{13}$CO ($B_0= 5.51 \times
10^{10}$ s$^{-1}$) and $Z$ is the partition function which is given by
\begin{equation}\label{eq:3}
Z=\sum^{\infty}_{J=0} (2J+1)e^\frac{-h B_0 (J+1)}{K T_{\rm ex}}.
\end{equation}


The partition function can be evaluated explicitly as a sum, but
\citet{Penzias1975} pointed out that for temperatures $T \gg hB_0/K$,
the partition function can be approximated by a definite integral,
which has value $kT/hB_0$.  This form for the partition function of a
rigid rotor molecule is almost universally employed, but it does
contribute a small error at the relatively low temperatures of dark
clouds. Specifically, the integral approximation always yields a value
of $Z$ which is smaller than the correct value.  Calculating $Z$
explicitly shows that this quantity is underestimated by a factor of
$\sim$1.1 in the range between 8 K to 10K.  Note that to evaluate
Equation~(\ref{eq:3}) we assume LTE (i.e. constant excitation
temperature) which might not hold for high--$J$ transitions. The error
due to this approximation is, however, very small. For example, for
$T_{\rm ex}$=10\,K, only 7\% of the populated states is at $J=3$
or higher.


We can calculate the column density of $^{13}$CO from
Equation~(\ref{eq:9}) determining the excitation temperature $T_{\rm
ex}$ and the $^{13}$CO optical depth from $^{12}$CO and $^{13}$CO
observations.   To estimate $T_{\rm ex}$ we assume that the
$^{12}$CO line is optically thick ($\tau \gg 1$) in
Equation~(\ref{eq:14}). This results in

\begin{equation}\label{eq:5}
T_\mathrm{\rm ex}=\frac{5.53}{ 
\ln \left ( 
1+\frac{5.53}{  T^{12}_\mathrm{mb,c}+0.83}
\right )},
\end{equation}
where $  T^{12}_\mathrm{mb,c}$ is the
peak corrected main-beam brightness temperature of $^{12}$CO. The excitation
temperature in Mask 2 ranges from 4 to 19\,K with a mean value of
9.7\,K and standard deviation of 1.2\,K.

Also from Equation~(\ref{eq:14}), the optical depth as a function of
velocity  of the $^{13}$CO $J = 1 \to 0$ line is obtained from the
main-beam brightness temperature using

 \begin{equation}\label{eq:4}
 \tau^{13}(v)=
 -\ln \left [  1 - \frac{  T^{13}_\mathrm{mb,c}(v)}{5.29} 
 \left ( 
 \left[
 e^{5.29/T_\mathrm{\rm ex}}  -1 
 \right]^{-1}-0.16 
 \right )^{-1} 
 \right], 
 \end{equation}
where $ T^{13}_\mathrm{mb,c}$ is the peak corrected main-beam
brightness temperature of $^{13}$CO.  We use this expression in
Equation~(\ref{cf1}) to determine opacity correction factor. We
evaluate the integrals in Equation~(\ref{cf1}) numerically. The
correction factor ranges from 1 to $\sim$4 with a mean value of 1.3
and standard deviation of 0.2.  The $^{13}$CO column density is
transformed to $^{12}$CO column density assuming a $^{12}$CO/$^{13}$CO
isotope ratio of 69 \citep{Wilson1999}, which should apply for the
well--shielded material in Mask\,2.


\subsubsection{Correction for Temperature Gradients along the Line of Sight}
\label{sec:corr-temp-grad}

In the derivation of the CO column density and its opacity correction
we made the assumption that the gas is isothermal. But observations
suggest the existence of core-to-edge temperature differences in
molecular clouds \citep[e.g.][]{Evans2001} which can be found even in
regions of only moderate radiation field intensity. Therefore the
presence of temperature gradients might affect our opacity correction.

We used the radiative transfer code RATRAN \citep{Hogerheijde2000} to
study the effects of temperature gradients on the determination of
$N({\rm CO})$. The modeling is described in the
Appendix~\ref{sec:radi-transf-model}.  We found that using $^{12}$CO
to determine the excitation temperature of the CO gas  gives the
correct temperature only at low column densities while the temperature is
overestimated for larger column densities.  This produces an
underestimate of the $^{13}$CO opacity which in turn affects the
opacity correction of $N({\rm CO})$. This results in an
underestimation of $N({\rm CO})$.  We derived a correction for this
effect (Equation~[\ref{eq:11}]) which is applied to the data.

\begin{figure*}[p]
\centering
\includegraphics[width=0.8\textwidth,angle=0]{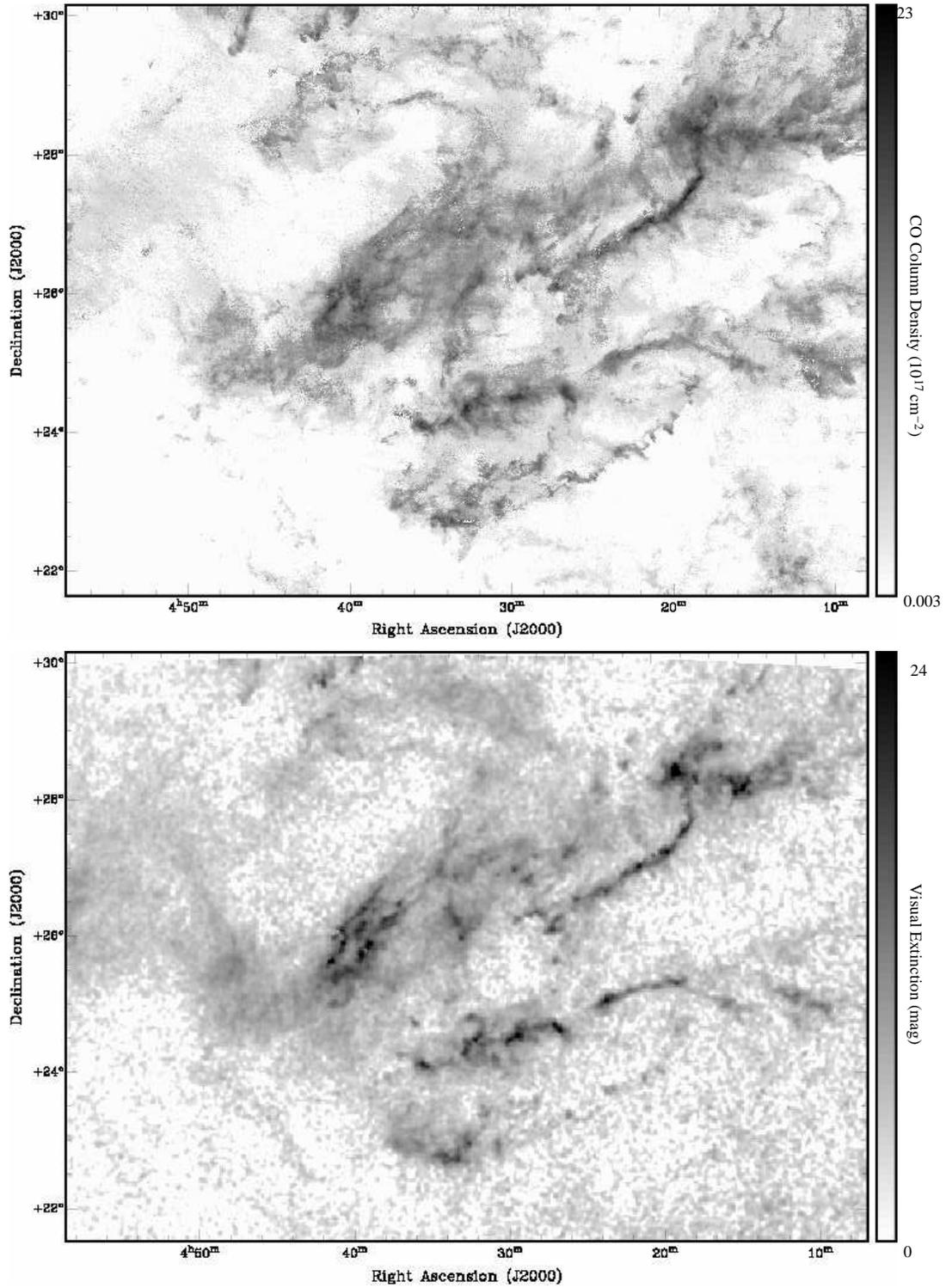}
\caption{Maps of the CO column density ({\it upper panel}) and visual
 extinction ({\it lower panel}) in  the Taurus Molecular cloud. The
 gray-scale in the $N({\rm CO})$ and $A_{\rm V}$ maps is expressed as
 the square root of the CO column density and of the visual
 extinction, respectively. The angular resolution of the data in the
 figure is 40\arcsec\,for $N({\rm CO})$ and 200\arcsec\,for $A_{\rm
 V}$.}\label{fig:maps_all}
\end{figure*} 

\subsection{CO Column Density in Mask 1}
\label{sec:total-co-column}


The column density of CO in molecular clouds is commonly determined
from observations of $^{12}$CO and $^{13}$CO with the assumption of
Local Thermodynamic Equilibrium (LTE), as discussed in the previous
section.  The lower limit of $N({\rm CO})$ that can be determined is
therefore set by the detection limit of the $^{13}$CO $J = 1 \to 0$
line.  For large maps, however, it is possible to determine $N({\rm
CO})$ in regions where only $^{12}$CO is detected in individual pixels
by using the statistical approach presented by
\citet{Goldsmith2008}. In the following we use this approach to
determine the column density of CO in Mask 1.

We compute the excitation temperature from the $^{12}$CO peak
intensities for all positions in Mask 1 assuming that the emission is
optically thick.  The Mask 1 data is then binned by excitation
temperature (in 1\,K bins), and the $^{13}$CO data for all positions
within each bin averaged together.  In all bins we get a very
significant detection of $^{13}$CO from the bin average.  Thus, we
have the excitation temperature and the observed ratio of integrated
intensities ($^{12}$CO/$^{13}$CO) in each 1\,K bin. Since positions in
Mask 1 are distributed in the periphery of high extinction regions, it
is reasonable to assume that the gas volume density in this region is
modest, and thus LTE does not necessarily apply, as thermalization
would imply an unreasonably low gas temperature at the cloud edges.
We therefore assume that $^{12}$CO is sub-thermally excited and that
the gas has a kinetic temperature of 15\,K.  We use the RADEX program
\citep{vanderTak2007}, using the LVG approximation, and the collision
cross sections from the Leiden Atomic and Molecular Database
\citep[LAMDA;][]{Schoeier2005}, to compute line intensities.  The free
parameters in the modeling are temperature ($T$), density ($n$), CO
column density per unit line width ($N({\rm CO})/\delta v$), and the
$^{12}$CO/$^{13}$CO abundance ratio ($R$).  Since the excitation is
determined by both density and the amount of trapping ($N/\delta v$),
there is a family of $n-N({\rm CO})/\delta v$ parameters that give the
same excitation temperature.  The other information we have is the
$^{13}$CO integrated intensity for the average spectrum in each bin.
Thus the choice of $n$, $N({\rm CO})/\delta v$ and $R$ must reproduce
the excitation temperature and the observed $^{12}$CO/$^{13}$CO ratio.
Solutions also must have an optical depth in the $^{12}$CO $J =1 \to
0$ of at least 3, to be consistent with the assumption that this
isotopologue is optically thick.  This is the same method used in
\citet{Goldsmith2008}, although this time we used the RADEX program
and the updated cross-sections from LAMDA.

  In fact, at low excitation temperature the data can only be fit
if the CO is strongly fractionated.  At high excitation temperature we
believe that the CO is unlikely to be fractionated, and thus, $R$ must
vary with excitation temperature.  We chose solutions for Mask 1 that
produced both a monotonically decreasing $R$ with decreasing
excitation temperature and a smoothly decreasing column density with
decreasing excitation temperature.  The solutions are given in
Table\,1 and shown Figure~\ref{fig:ron_results}.  The uncertainty
resulting from the assumption of a fixed kinetic temperature and from
choosing the best value for $R$ is about a factor of 2 in $N({\rm
CO})$ \citep{Goldsmith2008}.


To obtain $N({\rm CO})$ per unit line width for a given value of the
excitation temperature we have used a non-linear fit to the data, and
obtained the fitted function:

\begin{equation} 
     \left (  \frac{N({\rm CO})}{\rm cm^{-2}} \right ) \left ( \frac{\delta {\it v}}{\rm km\,s^{-1}}\right )^{-1} = 6.5\times10^{13} \left ( \frac{T_{\rm ex}}{{\rm K}} \right )^{2.7}.
\end{equation} 
 
We multiply by the observed FWHM line width to determine the total CO
column density. The upper panel in Figure~\ref{fig:ron_results} shows
 $N({\rm CO})/\delta v$ as a function of  $T_{\rm ex}$.

\subsection{CO Column Density in Mask 0}
\label{sec:co-column-densities_mask0}

To determine the carbon monoxide column density in regions where
neither $^{12}$CO nor $^{13}$CO were detected, we average nearly
10$^{6}$ spectra to obtain a single $^{12}$CO and $^{13}$CO spectra.
From the averaged spectra we obtain a $^{12}$CO/$^{13}$CO integrated
intensity ratio of $\simeq$17.  We need a relatively low $R$ to
reproduce such a low observed value.  Values of $R = 25$ or larger
cannot reproduce the observed isotopic ratio and still produce
$^{12}$CO emission below the detection threshold. Choosing $R = 20$
and a gas kinetic temperature of 15\,K, we fit the observed ratio
with $n = 100$\,cm$^{-3}$ and $N({\rm CO})= 3\times10^{15}$ cm$^{-2}$.
This gives rise to a $^{12}$CO intensity of 0.7 K, below the
detection threshold, however much stronger than the Mask 0 average of
only 0.18 K.  Thus, much of Mask 0 must not contribute to the CO
emission.  In fact, only 26\% of the Mask 0 area can have the
properties summarized above, producing significant CO emission.
Therefore, the average column density\footnote{Note that the estimate
of the CO column density in Mask\,0 by \citet{Goldsmith2008} did not
include the $\sim$26\% filling factor we derived here and in
consequence overestimated the CO column density in this region.}
throughout Mask 0 is $7.8\times10^{14}$\,cm$^{-2}$.

Another option is to model the average spectra of $^{12}$CO and
$^{13}$CO matching both the ratio and intensity.  Since now, our goal
is to produce CO emission with intensity 0.18\,K, both $^{12}$CO and
$^{13}$CO will be optically thin.  Therefore we need an $R$ that is
equal to the observed ratio.  For $R=18$, a solution with $n =
100$\,cm$^{-3}$, $\delta v = 1$\,km\,s$^{-1}$, and $N({\rm CO})$ =
7.3$\times10^{14}$\,cm$^{-2}$ fits both the $^{12}$CO and $^{13}$CO
average spectra for Mask 0.  Note that this is very similar to the
average solution (with a slightly larger $R$) that assumes that
$\sim$26\% of the area has column density 3$\times 10^{15}$ cm$^{-2}$
and the rest 0. Thus for a density of 100 cm$^{-3}$, the average CO
column density must be about 7.8$\times 10^{14}$\,cm$^{-2}$ in either
model.  Of course, if we picked a different density we would get a
slightly different column density.   As mentioned above, the
uncertainty is $N({\rm CO})$ is about a factor of 2.

Note that the effective area of CO emission is uniformly spread over
Mask\,0.  We subdivided the $^{12}$CO data cube in the Mask\,0 region
in an uniform grid with each bin containing about 10$^4$ pixels. After
averaging the spectra in each bin we find significant $^{12}$CO
emission in 95\% of them.

\begin{figure}[t]
\centering
\includegraphics[width=0.48\textwidth,angle=0]{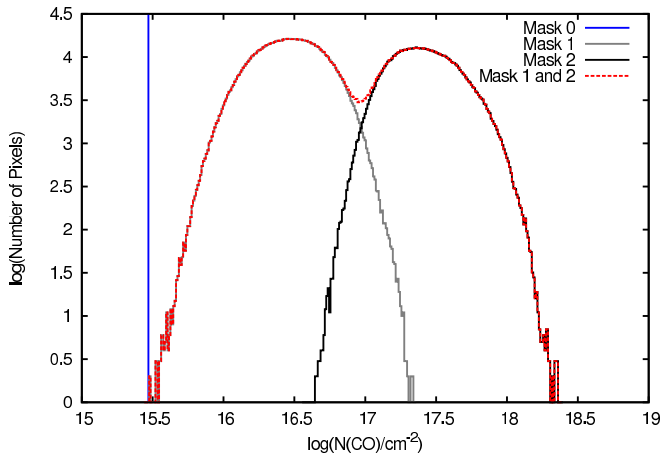}
\caption{Histogram of the $^{12}$CO column density distributions in
the Mask 0, 1, and 2 regions mapped in Taurus. The Mask 0 is indicated
by a vertical line at $N({\rm CO})=3 \times 10^{15}$\,cm$^{-2}$ which
represents the column density in the CO--emitting region (26\% of the
area of Mask 0; see Section~\ref{sec:co-column-densities_mask0}).
Note that we have not yet corrected $N({\rm CO})$ in Mask\,2 for
the effect of temperature gradients in the opacity
correction.}\label{fig:NCO_hist}
\end{figure} 

\section{Comparison between A$_{\rm V}$ and $N(^{12}{\rm CO})$ }
\label{sec:comp-betw-a_v}

\begin{figure}[t]
  \includegraphics[width=0.5\textwidth,angle=0]{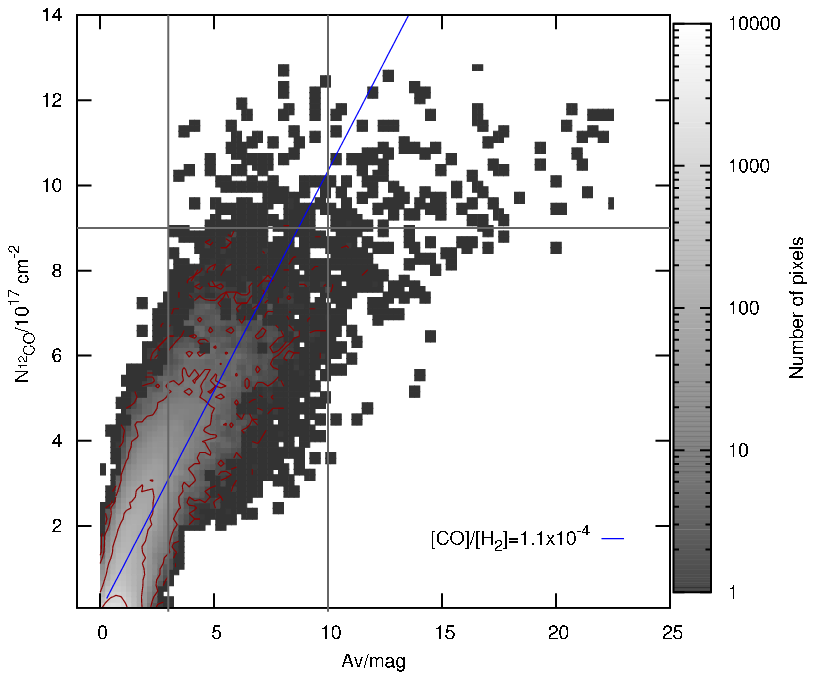}
     \caption{Comparison between the visual extinction derived from
2MASS stellar colors and the $^{12}$CO column density derived from
$^{13}$CO and $^{12}$CO observations in Taurus. The dark blue line
represents the $^{12}$CO column density derived from $A_{\rm V}$
assuming $N({\rm H}_2)/A_\mathrm{V} = 9.4\times10^{20} {\rm
cm}^{-2}\,{\rm mag}^{-1}$ \citep{Bohlin1978} and  
a [CO]/[H$_2$] abundance ratio of
$1.1\times10^{-4}$. The gray scale represents the number of pixels of
a given value in the parameter space and is logarithmic in the number
of pixels.  The red contours are 2,10,100, and 1000 pixels.  Each
pixel has a size of 100\arcsec\,or 0.07\,pc at a distance of 140
pc.}\label{fig:av_nh2_all}
\end{figure}


\begin{figure}
\includegraphics[width=0.47\textwidth,angle=0]{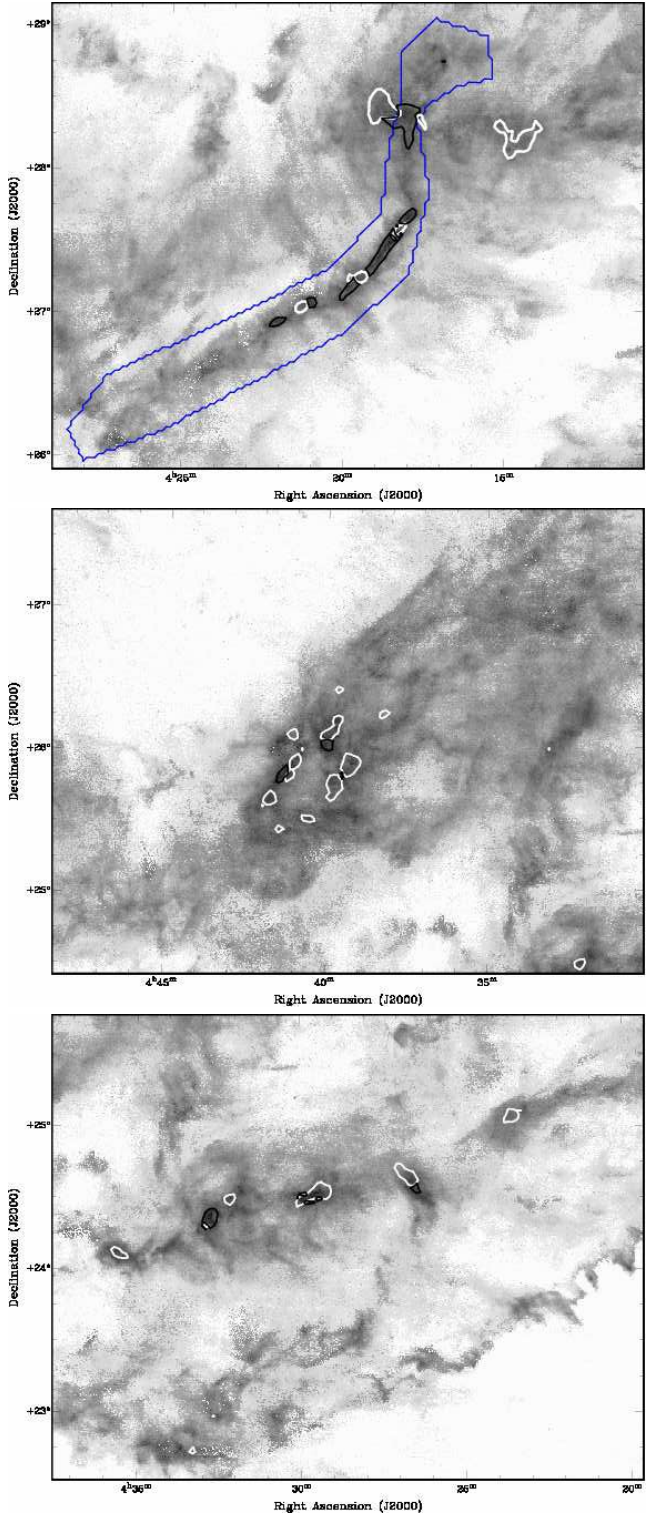}
\caption{$N({\rm CO})$ maps of the B213--L1495 ({\it top}), Heiles's
cloud 2 ({\it middle}), and B18--L1536 ({\it bottom}) regions. The
white contours denote regions with $A_{\rm V} > 10$ mag, while the
black contours denote regions with $A_{\rm V} < 10$ mag and $N({\rm
CO}) > 9 \times 10^{17}$\,cm$^{-2}$ (see
Figure~\ref{fig:av_nh2_all}). The blue contour outlines approximately
the B213--L1457 filament.  }\label{fig:zoom}
\end{figure}

\begin{figure*}[t]
  \includegraphics[width=1\textwidth,angle=0]{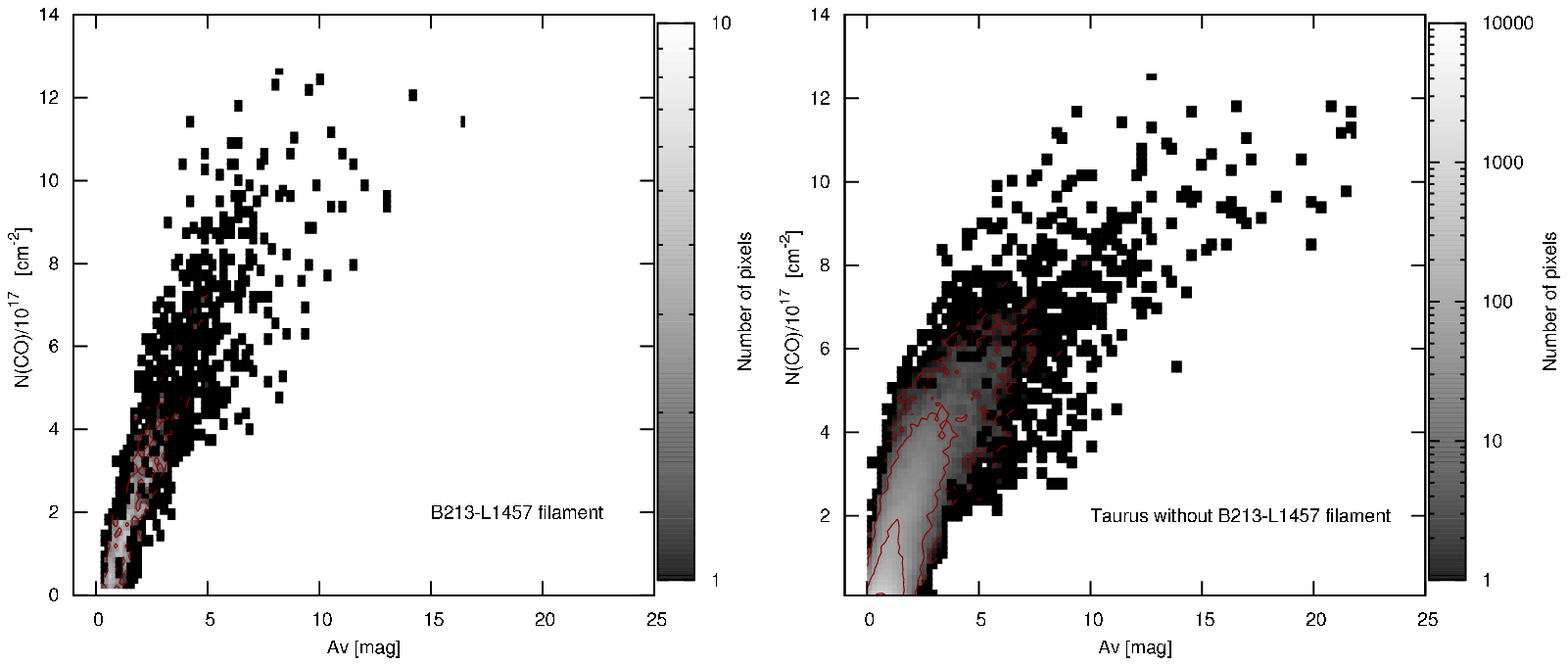}
     \caption{Pixel--by--pixel comparison between $A_{\rm V}$ and
     $N({\rm CO})$ in the B213--L1457 filament ({\it left}) and the
     entire Taurus molecular cloud without this filament ({\it
     right}).  }\label{fig:filament}
\end{figure*}


In order to test our estimate of $N({\rm CO})$ and assess whether it
is a good tracer of $N({\rm H}_2)$, we compare Mask\,1 and 2 in our CO
column density map of Taurus with a dust extinction map derived from
2MASS stellar colors.  Maps of these quantities are shown in
Figure~\ref{fig:maps_all}.  We also show in Figure~\ref{fig:NCO_hist}
a histogram of the $^{12}$CO column density distributions in the Mask
0, 1, and 2 regions mapped in Taurus.  The derivation of the dust
extinction map is described in Appendix~\ref{sec:extinction-map}. The
resolution of the map is 200\arcsec\,(0.14\,pc at a distance of
140\,pc) with a pixel spacing of 100\arcsec. For the comparison, we
have convolved and re-gridded the CO column density map in order to
match this resolution and pixel spacing.

\subsection{Large $N(^{12}{\rm CO})$  Column Densities}
\label{sec:large-n12rm-co}

 We show in Figure~\ref{fig:av_nh2_all} a pixel-by-pixel comparison
between visual extinction and $^{12}$CO column density. The visual
extinction and $N({\rm CO})$ are linearly correlated up to about
$A_{\rm V} \simeq 10$ mag. For larger visual extinctions $N({\rm CO})$
is largely uncorrelated with the value of $A_{\rm V}$.  In the range
$3 < A_{\rm V} < 10$\,mag, for a given value of $A_{\rm V}$, the mean
value of $N({\rm CO})$ is roughly that expected for a [CO]/[H$_2$]
relative abundance of $\sim$10$^{-4}$ which is expected for shielded
regions \citep{Solomon1972,Herbst1973}.  Some pixels, however, have CO column densities that
suggest a relative abundance that is reduced by up to a factor of
$\sim$3.  In the plot we show lines defining regions containing pixels
with $A_{\rm V} > 10$\,mag and with $3< A_{\rm V} < 10$ mag and
$N({\rm CO}) > 9 \times 10^{17}$\,cm$^{-2}$.  In Figure~\ref{fig:zoom}
we show the spatial distribution of these pixels in $N({\rm CO})$ maps
of the B213-L1457, Heiles's cloud 2, and B18-L1536 regions. White
contours correspond to the pixels with $A_{\rm V} > 10$\,mag and black
contours to pixels with $3< A_{\rm V} < 10$ mag and $N({\rm CO}) > 9
\times 10^{17}$\,cm$^{-2}$.  Regions with $A_{\rm V} > 10$\,mag are
compact and they likely correspond to the center of dense cores. The
largest values of $N({\rm CO})$, however, are not always spatially
correlated with such regions.  We notice that large $N({\rm CO})$ in
the $A_{\rm V}= 3-10$\,mag range are mostly located in the B213--L1457
filament.  We study the relation between $A_{\rm V}$ and $N({\rm CO})$
in this filament by applying a mask to isolate this region (see marked
region in Figure~\ref{fig:zoom}).  We show the relation between
$A_{\rm V}$ and $N({\rm CO})$ in the B213--L1457 filament in the left
hand panel of Figure~\ref{fig:filament}.  We also show this relation
for the entire Taurus molecular cloud excluding this filament in the
right hand panel.  Visual extinction and CO column density are
linearly correlated in the B213--L1457 filament with the exception of
a few pixels that are located in dense cores (Cores 3, 6 and 7 in
Table\,2). Without the filament the $N({\rm CO})$/$A_{\rm V}$ relation
is linear only up to $\sim$4 magnitudes of extinction.  In
Section~\ref{sec:co-depletion-1} we will see that the deviation from a
linear $N({\rm CO})$/$A_{\rm V}$ relation is mostly due to depletion
of CO molecules onto dust grains.  Depletion starts to be noticeable
for $A_{\rm V} \geq 4$\,mag.  Therefore, pixels on the B213--L1457
filament appear to show no signatures of depletion.  This can be due
either to the filament being chemically young in contrast with the
rest of Taurus, or to the volume densities being low enough that
desorption processes dominate over those of adsorption.  If the latter
case applies, and assuming a volume density of $n({\rm
H}_2)=10^{3}$\,cm$^{-3}$ (low enough to not show significant CO
depletion but still larger than the critical density of the $^{13}$CO
$J = 1\to 0$ line), this filament would need to be extended along the
line-of-sight by 0.9--3\,pc for $3 < A_{\rm V} < 10$\,mag. This length
is much larger than the projected thickness of the B213--L1495
filament of $\sim$0.2\,pc but comparable to its length of $\sim$7\,pc.
We will study the nature of this filament in a separate paper.

Considering only regions with $A_{\rm V} <10$\,mag and $N({\rm CO}) >
10^{17}$\,cm$^{-2}$ (see Section~\ref{sec:low-n12rm-co}) we fit a
straight line to the data in Figure~\ref{fig:av_nh2_all} to derive the
[CO]/[H$_2$] relative abundance in Mask\,2. A least squares fit
results in $N({\rm CO})/{\rm
cm}^{-2}=(1.01\pm0.008)\times10^{17}A_{\rm V}/{\rm mag}$.  
Assuming that all hydrogen is in molecular form we can write the ratio
between H$_2$ column density and color excess observed by
\citet{Bohlin1978} as $N({\rm H}_2)$/E$_{\rm
B-V}$=2.9$\times10^{21}$\,cm$^{-2}$ mag$^{-1}$. We combine this
relation with the ratio of total to selective extinction $R_{\rm
V}=A_{\rm V}/E_{\rm B-V}\simeq3.1$ \citep[e.g.][]{Whittet1992} to
obtain $N({\rm H}_2)/A_\mathrm{V} = 9.4\times10^{20} {\rm
22cm}^{-2}\,{\rm mag}^{-1}$. Combining the $N({\rm H}_2)$/$A_{\rm V}$
relation with our fit to the data, we obtain a [CO]/[H$_2$] relative
abundance of 1.1$\times10^{-4}$.  Note that, as discussed in
Appendix~\ref{sec:extinction-map}, grain growth would increase the
value of $R_{\rm V}$ up to $\sim$4.5 in dense regions
\citep{Whittet2001}. Due to this effect, we estimate that the derived
$A_{\rm V}$ would increase up to 20\% for $A_{\rm V}\leq$10\,mag. This
would reduce the $N({\rm H}_2)$/$A_{\rm V}$ conversion but also
increase the $A_{\rm V}$/$N({\rm CO})$ ratio.  Thus the derived
[CO]/[H$_2$] abundance is not significantly affected.

\begin{figure}[t]
\includegraphics[width=0.5\textwidth,angle=0]{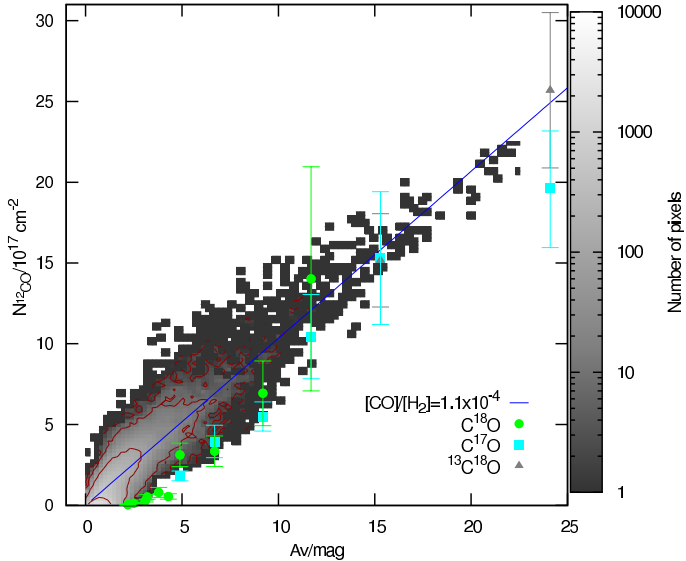}
\caption{The same as Figure~\ref{fig:av_nh2_all} but including the
 estimated column density of CO and CO$_2$ ices. For comparison we
 show the relation between visual extinction and $N({\rm CO})$ derived
 from observations of rare isotopic species by \citet{FLW1982} (see
 Appendix~\ref{sec:radi-transf-model}) which also include the
 contribution for CO and CO$_2$ ices.  }\label{fig:nco_av_corr_ice}
\end{figure}

\subsubsection{CO depletion}
\label{sec:co-depletion-1}


The flattening of the $A_{\rm V}$--$N({\rm CO})$ relation for $A_{\rm
V} > 10$\,mag could be due to CO depletion onto dust grains.  This is
supported by observations of the pre-stellar core B68 by
\citet{Bergin2002} which show a linear increase in the optically thin
C$^{18}$O and C$^{17}$O intensity as a function of $A_{\rm V}$ up to
$\sim$7\,mag, after which the there is a turnover in the intensity of
these molecules.  This is similar to what we see in
Figure~\ref{fig:av_nh2_all}.  Note, however, $A_{\rm V}$ alone is not
the sole parameter determining CO freeze-out, since this process also
depends on density and timescale \citep[e.g.][]{Bergin1997}.

Following \citet{Whittet2010}, we test the possibility
that effects of CO depletion are present in our observations of the
Taurus molecular cloud by accounting for the column of CO observed to
be in the form of ice on the dust grains.  \citet{Whittet2007}
measured the column density of CO and CO$_2$ ices\footnote{It is
predicted that oxidation reactions involving the CO molecules depleted
from the gas--phase can produce substantial amounts of CO$_2$ in the
surface of dust grains \citep{Tielens1982,Ruffle2001,Roser2001}. Since
the timescale of these his reactions are short compared with the
cloud's lifetime, we need to include CO$_2$ in order to account for
the amount of CO frozen into dust grains along the line--of--sight. }
toward a sample of stars located behind the Taurus molecular
cloud. They find that the column densities are related to the visual
extinction as

 \begin{equation}\label{eq:12}
\frac{N({\rm CO})_{\rm ice}}{10^{17}[{\rm cm}^{-2}]}=0.4(A_{\rm V}-6.7),\,\,\,A_{\rm V} > 6.7\,{\rm mag},
 \end{equation}
and
 \begin{equation}\label{eq:13}
\frac{N({\rm CO}_{2})_{\rm ice}}{10^{17}[{\rm cm}^{-2}]}=0.252(A_{\rm V}-4.0),\,\,\,A_{\rm V}> 4.0\,{\rm mag}.
 \end{equation}
We assume that the total column of CO frozen onto dust grains is given by

 \begin{equation}\label{eq:7}
N({\rm CO})^{\rm total}_{\rm ice}=N({\rm CO})_{\rm ice}+N({\rm CO}_{2})_{\rm ice}. 
 \end{equation}
Thus, for a given $A_{\rm V}$ the total CO column density is given by

\begin{equation}\label{eq:10}
N({\rm CO})^{\rm total}=N({\rm CO})_{\rm gas-phase}+N({\rm CO})^{\rm total}_{\rm ice}.
\end{equation}
We can combine our determination of the column density of gas-phase CO
with that of CO ices to plot the total $N({\rm CO})$ as a function of
$A_{\rm V}$. The result is shown in
Figure~\ref{fig:nco_av_corr_ice}. The visual extinction and $N({\rm
CO})^{\rm total}$ are linearly correlated over the entire range
covered by our data, extending up to $A_{\rm V}=23$\,mag. This result
confirms that depletion is the origin of the deficit of gas-phase CO
seen in Figure~\ref{fig:av_nh2_all}.

\begin{figure}[t]
\includegraphics[width=0.48\textwidth,angle=0]{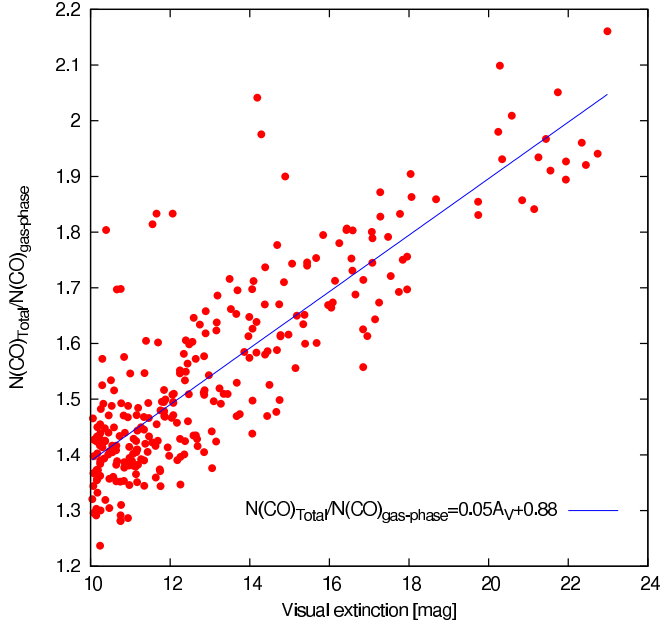}
\caption{Ratio of $N({\rm CO})_{\rm total}$ to $N({\rm CO})_{\rm
gas-phase}$ plotted as a function of $A_{\rm V}$ for the high
extinction portion of the Taurus molecular cloud. The line represents
our fit to the data. }\label{fig:map}
\end{figure}

In Figure~\ref{fig:map} we show the ratio of $N({\rm CO})^{\rm total}$
to $N({\rm CO})_{\rm gas-phase}$ as a function of $A_{{\rm V}}$, for
$A_{\rm V}$ greater than 10.  The drop in the relative abundance of
gas-phase CO from our observations is at most a factor of
$\sim$2. This is in agreement with previous determinations of the
depletion along the line of sight in molecular clouds
\citep{Kramer1999,Chiar1995}.

\subsubsection{CO Depletion Age}
\label{sec:co-depletion}

In this Section we estimate the CO depletion age (i.e. the time needed
for CO molecules to deplete onto dust grains to the observed levels)
in dense regions in the Taurus Molecular Cloud. We selected a sample
of 13 cores that have peak visual extinction larger than 10\,mag 
and that $A_{\rm V}$ at the edges drops below $\sim$0.9\,mag (3 times the
uncertainty in the determination of $A_{\rm V}$). The cores are located in the L1495 and
B18--L1536 regions (Figure~\ref{fig:zoom}).  Unfortunately, we were
not able to identify individual cores in Heiles's Cloud 2 due to
blending.

We first determine the H$_{2}$ volume density structure of our
selected cores.  \citet{Dapp2009} proposed using the \citet{King1962}
density profile,

\begin{equation}
n(r)=\begin{cases}
 n_c a^2 / (r^2 + a^2) & r \leq R \\
0   &   r > R,
\end{cases}
\end{equation}
which is characterized by the central volume density $n_c$, a
truncation radius $R$, and by a central region of size $a$ with
approximately constant density.

The column density $N(x)$ at an offset from the core center $x$ can be
derived by integrating the volume density along a line of sight through
the sphere. Defining $N_c\equiv 2 a n_c \arctan(c)$ and $c=R/a$, the
column density can be written

\begin{deluxetable*}{lllcccccc}
\tabletypesize{\scriptsize} \centering \tablecolumns{9} \small
\tablewidth{0pt}
\tablecaption{Core Parameters}\label{sec:radi-transf-model-1}
\tablenum{2}
\tablehead{\colhead{Core ID} &  \colhead{$\alpha$(J2000)}   &\colhead{$\delta$(J2000)}&\colhead{$A_{{\rm V,}c}$} & \colhead{$a$} &   \colhead{Radius} & \colhead{$n_{c}({\rm H}_2)$}&\colhead{Mass}  & \colhead{Depletion Age}\\ 
\colhead{} & \colhead{} &\colhead{} &  \colhead{[mag]} & \colhead{[pc]} &   \colhead{[pc]} & \colhead{[$10^{4}$\,cm$^{-3}$]}&\colhead{[M$_\odot$]}  & \colhead{[$10^{5}$\,years]}}
\startdata
1  &04:13:51.63 & 28:13:18.6 & 22.4$\pm$0.5 & 0.10$\pm$0.004 & 2.01$\pm$0.30 & 2.2$\pm$0.11 & 307$\pm$102 &  6.3$\pm$0.3   \\
2  &04:17:13.52 & 28:20:03.8 & 10.7$\pm$0.3 & 0.19$\pm$0.021 & 0.54$\pm$0.12 & 0.7$\pm$0.09 & 56$\pm$43 &    3.4$\pm$1.5     \\
3  &04:18:05.13 & 27:34:01.6 & 12.3$\pm$1.2 & 0.16$\pm$0.054 & 0.32$\pm$0.18 & 1.0$\pm$0.40 & 29$\pm$63 &    1.3$\pm$3.1     \\
4  &04:18:27.84 & 28:27:16.3 & 24.2$\pm$0.4 & 0.13$\pm$0.005 & 1.27$\pm$0.08 & 1.9$\pm$0.07 & 258$\pm$52 &   3.8$\pm$0.2    \\
5 &04:18:45.66 & 25:18:0.4  & 9.4$\pm$0.2  & 0.09$\pm$0.005 & 2.00$\pm$0.93 & 1.1$\pm$0.07 & 110$\pm$81 &  10.9$\pm$0.8    \\
6 &04:19:14.99 & 27:14:36.4 & 14.3$\pm$0.6 & 0.12$\pm$0.010 & 0.89$\pm$0.15 & 1.3$\pm$0.12 & 93$\pm$47 &    3.1$\pm$0.6    \\
7  &04:21:08.46 & 27:02:03.2 & 15.2$\pm$0.3 & 0.08$\pm$0.003 & 1.12$\pm$0.08 & 1.9$\pm$0.07 & 90$\pm$19 &    2.9$\pm$0.2     \\
8  &04:23:33.84 & 25:03:01.6 & 14.4$\pm$0.3 &  0.11$\pm$0.004 & 0.93$\pm$0.07 & 1.4$\pm$0.06 & 94$\pm$23 &    5.1$\pm$0.3     \\
9  &04:26:39.29 & 24:37:07.9 & 15.6$\pm$0.5 & 0.09$\pm$0.006 & 1.48$\pm$0.23 & 1.7$\pm$0.12 & 143$\pm$58 &   2.3$\pm$0.3    \\
10  &04:29:20.71 & 24:32:35.6 & 17.2$\pm$0.4 & 0.13$\pm$0.006 & 2.48$\pm$0.39 & 1.3$\pm$0.07 & 371$\pm$127 &  4.5$\pm$0.3   \\
11 &04:32:09.32 & 24:28:39.0 & 16.0$\pm$0.5 & 0.09$\pm$0.006 & 3.50$\pm$2.41 & 1.7$\pm$0.13 & 347$\pm$347 & 3.2$\pm$0.4   \\
12 &04:33:16.62 & 22:42:59.6 & 12.2$\pm$0.5 & 0.08$\pm$0.007 & 1.66$\pm$0.64 & 1.5$\pm$0.14 & 110$\pm$82 &  6.3$\pm$0.7   \\
13  &04:35:34.29 & 24:06:18.2 & 12.5$\pm$0.3 & 0.11$\pm$0.006 & 1.64$\pm$0.35 & 1.2$\pm$0.07 & 145$\pm$65 &   2.0$\pm$0.4    \\
\enddata
\end{deluxetable*}
%
%

\begin{align}  
& N(x) = \frac{N_c}{\sqrt{1+(x/a)^2}}  \nonumber \\  
&\times \left [ \arctan(\sqrt{\frac{c^2-(x/a)^2}{1+(x/a)^2}})/\arctan(c)  \right ].
\end{align}
This column density profile can be fitted to the data. The three
parameters to fit are (1) the outer radius $R$, (2) the central column
density $N_c$ (which in our case is $A_{\rm V,c}$), and (3) the size
of the uniform density region $a$.

\begin{figure}
\includegraphics[width=0.455\textwidth,angle=0]{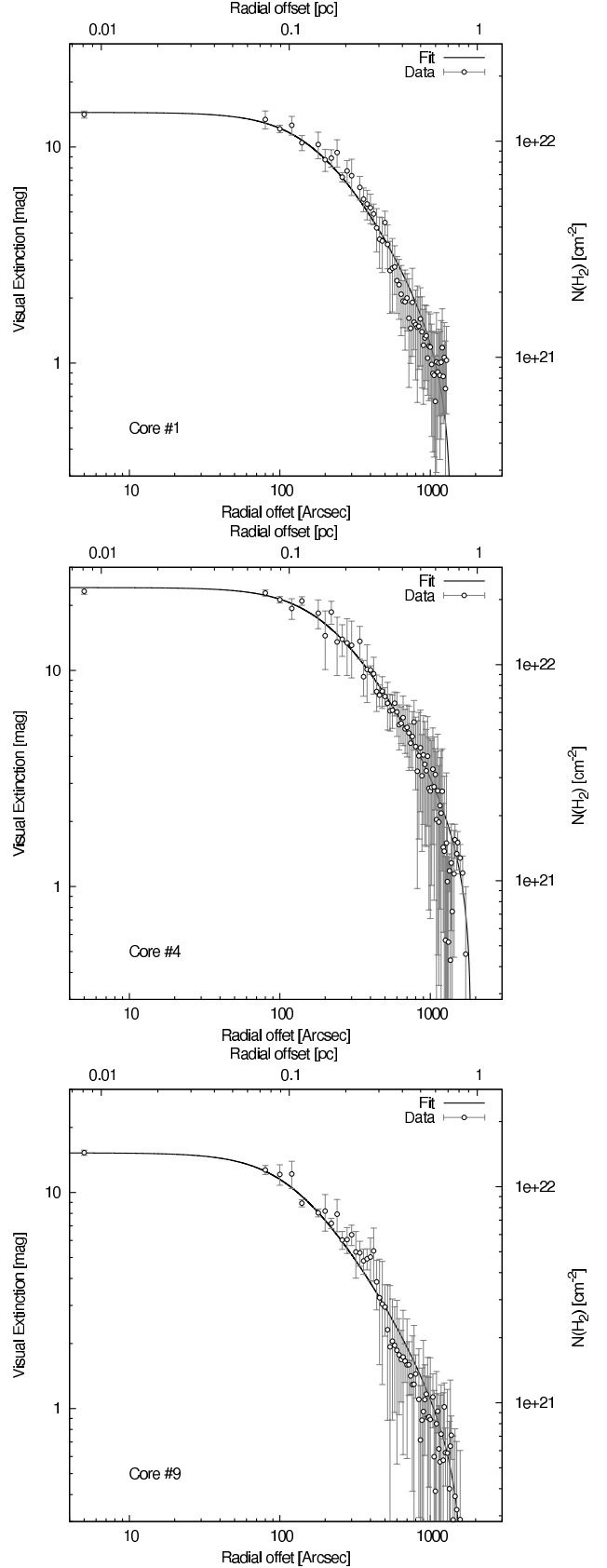}
\caption{Typical radial distributions of the visual
extinction in the selected sample of cores. The solid lines represent the corresponding fit. }\label{fig:fig}
\end{figure}

We obtain a column density profile for each core by fitting an
elliptical Gaussian to the data to obtain its central coordinates,
position angle, and major and minor axes. With this information we
average the data in concentric elliptical bins. Typical column density
profiles and fits to the data are shown in Figure~\ref{fig:fig}. We
give the derived parameters of the 13 cores we have analyzed in
Table\,2. We convert the visual extinction at the core center $A_{\rm
V,c}$ to H$_2$ column density assuming $N({\rm H}_2)/A_\mathrm{V} =
9.4\times10^{20} {\rm cm}^{-2}\,{\rm mag}^{-1}$. We use then the
definition of column density at the core center (see above) to
determine the central volume density $n_c({\rm H}_2)$ from the fitted
parameters.

With the H$_2$ volume density structure, we can derive the CO
depletion age for each core.  The time needed for CO molecules to
deplete to a specified degree onto dust grains is given by
\citep[e.g.][]{Bergin07},

\begin{equation}
t_{\rm depletion}=\left ( \frac{5\times10^{9}}{{\rm yr}} \right) \left( \frac{n({\rm H}_2)}{{\rm cm}^{-3}}  \right)^{-1} \ln(n_0/n_{\rm gas}),
\end{equation}
where $n_0$ is the total gas--phase density of CO before depletion
started and $n_{\rm gas}$ the gas-phase CO density at time $t_{\rm
depletion}$. Here we assumed  a sticking coefficient\footnote{ The
sticking coefficient is defined as how often a species will remain on
the grain upon impact \citep{Bergin07}.  } of unity
\citep{Bisschop2006} and that at the H$_{2}$ volume densities of
interest adsorption mechanisms dominate over those of desorption 
(we therefore assume that the desorption rate is zero).

To estimate $n_0/n_{\rm gas}$ we assume that CO depletion occurs only
in the flat  density region of a core, as for larger radii the volume
density drops rapidly. Then the total column density of CO
(gas--phase+ices) in this region is given by $N({\rm CO})^{\rm
flat}\simeq2an_c(1.1\times10^{-4})$. The gas-phase CO column density
in the flat density region of a core is given by $N^{\rm flat}_{\rm
gas-phase}({\rm CO})=N({\rm CO})^{\rm flat}-N({\rm CO})^{\rm
total}_{\rm ice}$, where $N({\rm CO})^{\rm total}_{\rm ice}$ can be
derived from Equation~(\ref{eq:10}). Assuming that the decrease in the
[CO]/[H$_2$] relative abundance in the flat region is fast and stays
constant toward the center of the core (models from
\citet{Tafalla2002} suggest an exponential decrease), then
$n_0/n_{\rm gas}\simeq N({\rm CO})^{\rm flat}/N^{\rm flat}_{\rm
gas-phase}({\rm CO})$. The derived CO depletion ages are listed in
Table\,2. Note that the fitted cores might not be fully resolved at the
resolution of our $A_{\rm V}$ map (200\arcsec\,or 0.14\,pc at a
distance of 140\,pc). Although $n_0/n_{\rm gas}$ is not very sensitive
to resolution, due to mass conservation, we might be underestimating
the density at the core center. Therefore, our estimates of the CO
depletion age might be considered as upper limits.

\begin{figure}
\includegraphics[width=0.48\textwidth,angle=0]{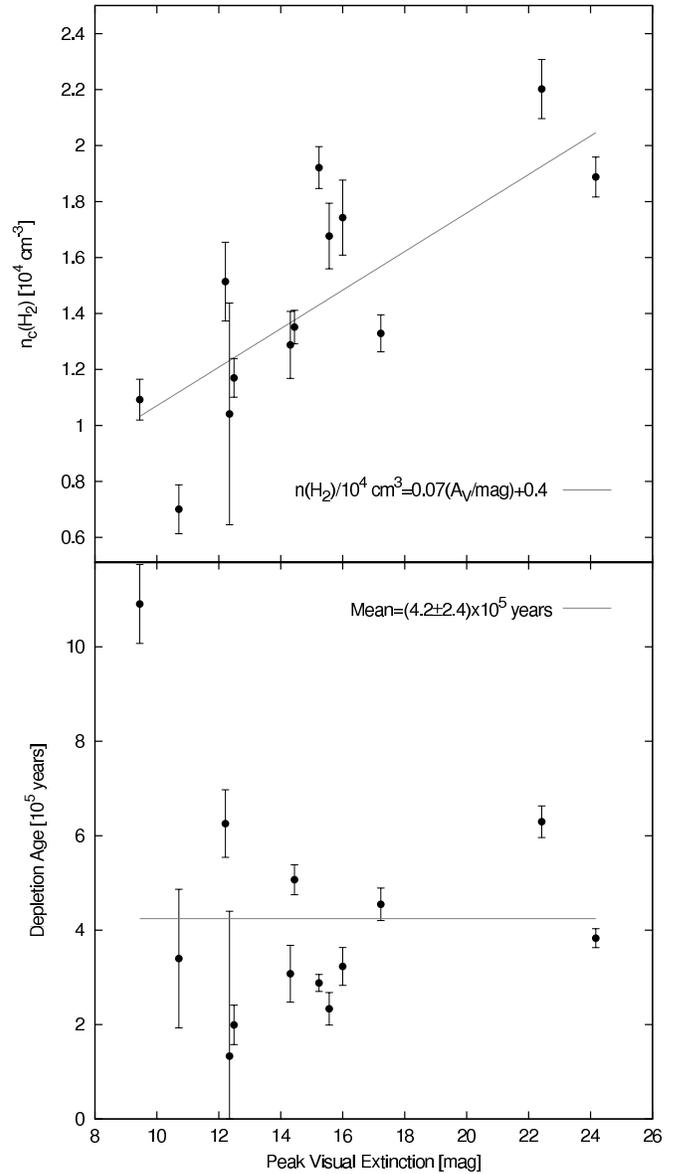}
\caption{({\it upper panel}) The central H$_2$ volume density as a function of
 the peak $A_{\rm V}$ for a sample of 13 cores in the Taurus molecular cloud. The
line represent a fit to the data. ({\it lower panel}) CO depletion age
as a function of  $A_{\rm V}$ for the sample of
cores. }\label{fig:plot_tdepletion}
\end{figure}

 In Figure~\ref{fig:plot_tdepletion} we show the central density and
the corresponding depletion age of the fitted cores as a function of
$A_{\rm V}$. The central volume density is well correlated with
$A_{\rm V}$ but varies only  over a small range: its mean value and standard
deviation are ($1.4\pm0.4$)$\times10^{4}$\,cm$^{-3}$. Still, the
moderate increase of $n({\rm H}_2)$ with $A_{\rm V}$ compensates for
the increase of $N({\rm CO})_{\rm total}/N({\rm CO})_{\rm gas-phase}$
with $A_{\rm V}$ to produce an almost constant depletion age. The mean
value and standard deviation of $t_{\rm depletion}$ are
$(4.2\pm2.4)\times10^{5}$\,years. This suggests that dense cores
attained their current central densities at a similar moment in the
history of the Taurus molecular cloud.

\subsection{Low $N(^{12}{\rm CO})$ column densities}
\label{sec:low-n12rm-co}

\begin{figure*}
\centering
  \includegraphics[width=0.75\textwidth,angle=0]{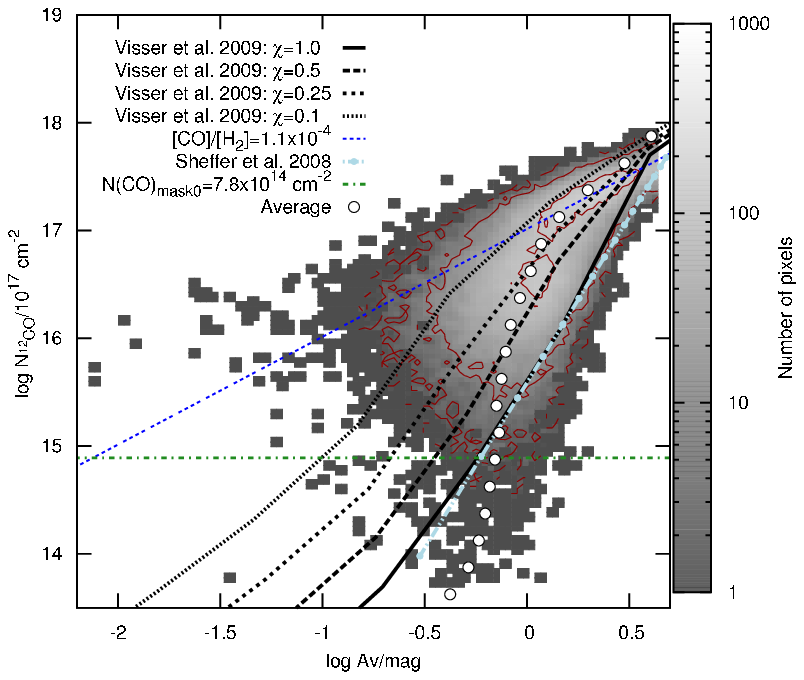}
     \caption{Comparison between the visual extinction derived from
2MASS stellar colors and the $^{12}$CO column density derived from
$^{13}$CO and $^{12}$CO  observations in Taurus for $A_{\rm V} < 5$\,mag. The blue
line represents the $^{12}$CO column density derived from $A_{\rm V}$
assuming $N({\rm H}_2)/A_\mathrm{V} = 9.4\times10^{20} {\rm cm}^{-2}\,{\rm
mag}^{-1}$ \citep{Bohlin1978} 
and a [CO]/[H$_2$] abundance ratio of $1.1\times10^{-4}$. The
gray scale represents the number of pixels of a given value in the
parameter space and is logarithmic in the number of pixels.  The red
contours are 2,10,100, and 1000 pixels.  The black lines represent
several models of selective CO photodissociation and fractionation
provided by Ruud Visser (see text).  The light blue line
represents the fit to the observations from \citet{Sheffer2008} toward
diffuse molecular Galactic lines-of-sight for log($N({\rm H})_2) \geq
20.4$. The horizontal line represents the average $N({\rm CO})$
derived in Mask 0. Each pixel has a size of 100\arcsec\,or 0.07\,pc at
a distance of 140 pc.}\label{fig:low_av}
\end{figure*}

\begin{figure}
\centering
  \includegraphics[width=0.49\textwidth,angle=0]{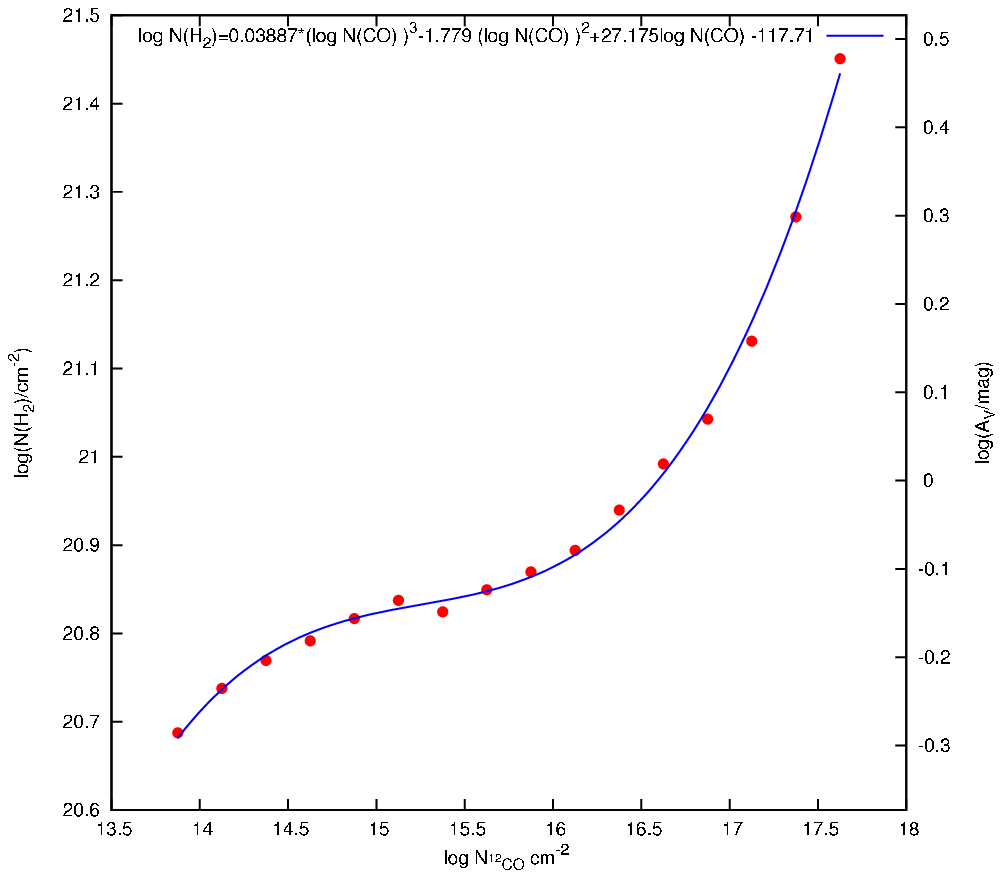}
     \caption{The average $N({\rm H}_2)$ and $A_{\rm V}$ as a function
     of $N({\rm CO})$ in Mask1. $N({\rm H}_2)$ is estimated from
     $A_{\rm V}$ assuming $N({\rm H}_2)/A_\mathrm{V} =
     9.4\times10^{20} {\rm cm}^{-2}\,{\rm mag}^{-1}$
     \citep{Bohlin1978}. }\label{fig:fit}
\end{figure}

In the following we compare the lowest values of the CO column density
in our Taurus survey with the visual extinction derived from 2MASS
stellar colors. In Figure~\ref{fig:low_av} we show a comparison
between $N({\rm CO})$ and $A_{\rm V}$ for values lower than 5
magnitudes of visual extinction. The figure includes CO column
densities for pixels located in Mask 1 and 2. We do not include pixels
in Mask 0 because its single value does not trace variations with
$A_{\rm V}$. Instead, we include a horizontal line indicating the
derived average CO column density in this Mask region. We show a
straight line (blue) that indicates $N({\rm CO})$ expected from a
abundance ratio [$^{12}$CO]/[H$_2$]=1.1$\times$10$^{-4}$
(Section~\ref{sec:large-n12rm-co}). The points indicate the
average $A_{\rm V}$ in a $N({\rm CO})$ bin. We present a fit to this
relation in Figure~\ref{fig:fit}.

The data are better described by a varying [$^{12}$CO]/[H$_2$]
abundance ratio than a fixed one. This might be caused by
photodissociation and fractionation of CO which can produce strong
variations in the CO abundances between UV-exposed and shielded
regions \citep{vanDishBlack88,Visser2009}.  To test this possibility
we include in the figure several models of these effects provided by
Ruud Visser (see \citealt{Visser2009} for details). They show the
relation between $A_{\rm V}$ and $N({\rm H}_2)$ for different values
of the FUV radiation field starting from $\chi$ = 1.0 to 0.1 (in units
of the mean interstellar radiation field derived by
\citealt{Draine78}).  All models have a kinetic temperature of 15\,K
and a total H volume density of 800\,cm$^{-3}$ which corresponds to
$n({\rm H}_2)\simeq 395\,{\rm cm}^{-3}$ assuming $n$(H\,{\sc
i})$=10$\,cm$^{-3}$. (This value of $n({\rm H}_2)$ is close to the
average in Mask\,1 of 375\,cm$^{-3}$.)  The observed relation between
$A_{\rm V}$ and $N({\rm CO})$ cannot be reproduced by a model with a
single value of $\chi$. This suggests that the gas have a range of
physical conditions.  Considering the average value of $A_{\rm V}$
within each bin covering a range in $N({\rm CO})$ of 0.25 dex, we see
that for an increasing value of the visual extinction, the FUV radiation
field is more and more attenuated so that we have a value of $N({\rm CO})$ that is
predicted by a model with reduced $\chi$.

%
%

We also include in Figure~\ref{fig:low_av} the fit to the observations
from \citet{Sheffer2008} toward diffuse molecular Galactic
lines--of--sight for log($N({\rm H}_2)) \geq 20.4$. The fit seems to
agree with the portion our data points that agree fairly well with the
model having $\chi=1.0$. Since \citet{Sheffer2008} observed diffuse
lines-of-sight, this suggests that a large fraction of the material in
the Taurus molecular cloud is shielded against the effect of the FUV
illumination.  This is supported by infrared observations in Taurus by
\citet{Flagey2009} that suggest that the strength of the FUV radiation
field is between $\chi= 0.3$ and 0.8.

\citet[][ see also \citealt {Federman1980}]{Sheffer2008} showed
empirical and theoretical evidence that the scatter in the $A_{\rm
V}-N({\rm CO})$ relation is due to variations of the ratio between the
total H volume density ($n^{\rm total}_{\rm H}=n_{\rm H}+2n_{\rm H_2}$) and the
strength of the FUV radiation field. The larger the volume density or
the weaker the strength of the FUV field the larger the abundance of
CO relative to H$_2$. Note that the scatter in the observations from
\citet{Sheffer2008} is much smaller than that shown in
Figure~\ref{fig:low_av}. This indicates that we are tracing a
wider range of physical conditions of the gas. The
excitation temperature of the gas observed by \citet{Sheffer2008} does
not show a large variation from $T_{\rm ex}$=5\,K while we observe
values between 4 and 15\,K.


In Figure~\ref{fig:low_av} we see that some regions can have large
[$^{12}$CO]/[H$_2$] abundance ratios but still have very small column
densities ($A_{\rm V}=0.1-0.5$\,mag).  This can be understood in terms
of a medium which is made of an ensemble of spatially unresolved dense
clumps embedded in a low density interclump medium \citep{Bensch06}.
 In this scenario, the contribution to the total column density
from dense clumps dominates over that from the tenuous inter--clump
medium.  Therefore the total column density is proportional to the
number of clumps along the line--of--sight.    
A low number of clumps along a line--of--sight would
give low column densities while in the interior of these dense clumps
CO is well shielded against FUV photons and therefore it can reach the
asymptotic value of the [CO]/[H$_2$] ratio characteristic of dark clouds.


\section{Discussion}
\label{sec:discussion}

\subsection{The mass of the Taurus Molecular Cloud}
\label{sec:mass-taur-molec}

In this section we estimate the mass of the Taurus Molecular Cloud
using the $N({\rm CO})$ and $A_{\rm V}$ maps. The masses derived for
Mask 0, 1, and 2 are listed in Table\,3. To derive the H$_2$ mass from
$N({\rm CO})$ we need to apply an appropriate [CO]/[H$_2$] relative
abundance for each mask. The simplest case is Mask 2 where we used the
asymptotic $^{12}$CO abundance of 1.1$\times$10$^{-4}$ (see
Section~\ref{sec:large-n12rm-co}). We corrected for saturation
including temperature gradients and for depletion in the mass
calculation from $N({\rm CO})$. These corrections amount to
$\sim$319\,M$_{\odot}$ (4\,M$_{\odot}$ from the saturation correction
and 315\,M$_{\odot}$ from the addition of the column density of
CO--ices).  For Mask 1 and 0, we use the fit to the relation between
$N({\rm H}_2)$ and $N({\rm CO})$ shown in Figure~\ref{fig:fit}.  As we
can see in Table\,3, the masses derived from $A_{\rm V}$ and $N({\rm
CO})$ are very similar. This confirms that $N({\rm CO})$ is a good
tracer of the bulk of the molecular gas mass if variations of the
[CO]/[H$_2$] abundance ratio are considered.

Most of the mass derived from $A_{\rm V}$ in Taurus is in Mask 2
($\sim$49\%). But a significant fraction of the total mass lies in
Mask 1 ($\sim$28\%) and Mask 0 ($\sim$23\%). This implies that mass
estimates that only consider regions where $^{13}$CO is detected
underestimate the total mass of the molecular gas by a factor of
$\sim$2.

We also estimate the masses of high--column density regions considered
by \citet{Goldsmith2008} that were previously defined by
\citet{Onishi1996}. In Table\,4 we list the masses derived from the
visual extinction as well as from $N({\rm CO})$. Again, both methods
give very similar masses.  These regions together represent 43\%
of the total mass in our map of Taurus, 32\% of the area, and 46\% of
the $^{12}$CO luminosity. This suggest that the mass and $^{12}$CO
luminosity are uniformly spread over the area of our Taurus map.

A commonly used method to derive the mass of molecular clouds when
only $^{12}$CO is available is the use of the empirically derived
CO--to--H$_2$ conversion factor ($X_{\rm CO} \equiv N({\rm
H}_2)/I_{\rm CO} \simeq M_{{\rm H}_2}/L_{\rm CO}$ ). Observations of
$\gamma$-rays indicate that this factor is
1.74$\times10^{20}$\,cm$^{-2}$(K km s$^{-1}$ pc$^{-2}$)$^{-1}$ or
$M({\rm M}_{\odot}$)=3.7$L_{\rm CO}$ (K Km s$^{-1}$ pc$^{2}$) in our Galaxy
\citep{Grenier2005}.  To estimate $X_{ \rm CO}$ in Mask 2, 1, and 0 we
calculate the $^{12}$CO luminosity ($L_{\rm CO}$) in these regions and
compare them with the mass derived from $A_{\rm V}$. We also calculate
$X_{\rm CO}$ from the average ratio of $N({\rm H}_2)$ (derived from
$A_{\rm V}$ ) to the CO integrated intensity $I_{\rm CO}$ for all
pixels in Mask\,1 and 2. For Mask\,0, we used the ratio of the average
$N({\rm H}_2)$ (derived from $A_{\rm V}$ ) to the average CO
integrated intensity obtaining after combining all pixels in this mask
region.  The resulting values are shown in Table\,3.  The table shows
that the difference in $X_{\rm CO}$ between Mask\,2 and Mask\,1 is
small considering that the [CO]/[H$_2$] relative abundance between
these regions can differ by up to two orders of magnitude.  The
derived values are close to that found in our Galaxy using
$\gamma$-ray observations. For Mask\,0, however, $X_{\rm CO}$ is about
an order of magnitude larger than in Mask\,1 and 2.

Finally we derive the surface density of Taurus by comparing the total
H$_2$ mass derived from $A_{\rm V}$ (15015\,M$_{\odot}$) and the total
area of the cloud (388\,pc$^2$). Again, we assumed that in Mask 0 the
CO--emitting region occupies  26\% of the area. The resulting surface density
is $\sim$39\,M$_{\odot}$\,pc$^{-2}$ which is very similar to the
median value of 42\,M$_{\odot}$\,pc$^{-2}$ derived from a large sample
of galactic molecular clouds by \citet{Heyer2008}.

\begin{deluxetable*}{lcccccccc}
\tabletypesize{\scriptsize}
\centering
\tablecolumns{9}
\small
\tablewidth{0pt}
\tablecaption{Properties of Different Mask Regions in Taurus}

\tablenum{3}
\tablehead{\colhead{Region} & \colhead{\# of Pixels$^{a}$} &\colhead{Mass from}   & \colhead{Mass from}   & \colhead{Area} & \colhead{$[{\rm CO}]/[{\rm H}_2]$} & \colhead{$L_{\rm CO}$} & \colhead{$X^{b}_{\rm CO}=N({\rm H}_2)$/$I_{\rm CO}$}  & \colhead{$X^{c}_{\rm CO}=M/L_{\rm CO}$}   \\  &  \colhead{}  & \colhead{$^{13}$CO and $^{12}$CO} & \colhead{$A_{\rm V}$}  &  \colhead{}  & \colhead{} & \colhead{}  & \colhead{} \\ &  \colhead{} &\colhead{[$M_\odot$]} & \colhead{[$M_\odot$]}  & \colhead{[pc$^{2}$]}   & \colhead{} & \colhead{[K km s$^{-1}$ pc$^{2}$]}   &  [cm$^{-2}$/(K km s$^{-1}$)] &  [$M_\odot$/(K km s$^{-1}$ pc$^{2}$)] }
\startdata
Mask 0 &52338& 3267  & 3454  & 63$^{d}$  &  1.2$\times10^{-6}$ &  130   & 1.2$\times10^{21}$  & 26 \\
Mask 1 &40101&  3942  & 4237   & 185 &  variable        &  1369  & 1.6$\times10^{20}$  & 3.1 \\
Mask 2 &30410& 7964  & 7412  & 140 & 1.1$\times10^{-4}$        &  1746  & 2.0$\times10^{20}$  & 4.2\\
Total  &122849& 15073 & 15103 &  388  &                         &  3245  & 2.3$\times10^{20}$ & 4.6   \\
\enddata
\tablenotetext{a}{At the 200\arcsec\,resolution of the $A_{\rm V}$ map.}
\tablenotetext{b}{Calculated from the average ratio of $N({\rm H}_2)$, derived from $A_{\rm V}$, to CO integrated intensity for each pixel.  }
\tablenotetext{c}{Total mass per unit of CO luminosity.}
\tablenotetext{d}{Effective area of CO emission based in the discussion about Mask\,0 in Section~\ref{sec:co-column-densities_mask0}.}
\end{deluxetable*}
%
%

%
%
%
%
%

\subsection{Column density  probability density function}
\label{sec:column-prob-dens}

\begin{deluxetable*}{lccccc}
\tabletypesize{\scriptsize}
\centering
\tablecolumns{5}
\small
\tablewidth{0pt}
\tablecaption{Mass of Different High Column Density  Regions in Taurus}
\tablenum{4}
\tablehead{\colhead{Region} & \colhead{\# of Pixels} & \colhead{Mass from}  & \colhead{Mass from}   & \colhead{Area}& \colhead{$L_{\rm CO}$} \\ \colhead{} & \colhead{} & \colhead{$^{13}$CO and $^{12}$CO}  & \colhead{$A_{\rm V}$}   & \colhead{} \\  &  & \colhead{[$M_\odot$]}  & \colhead{[$M_\odot$]}   & \colhead{[pc$^2$]} & \colhead{[K km s$^{-1}$ pc$^{2}$]}}
\startdata
L1495 &7523 & 1836 & 1545 & 35 & 461 \\
B213  &2880 & 723 & 640 & 13 & 155   \\
L1521 &4026 & 1084 & 1013 & 19 & 236 \\
HCL2  &3633 & 1303 & 1333 & 17 & 221 \\
L1498 &1050 & 213 & 170 & 5 & 39	    \\
L1506 &1478 & 262 & 278 & 7 & 68	    \\
B18   &3097 & 828 & 854 & 14 & 195   \\
L1536 &3230 & 474 & 579 & 15 & 134   \\
Total & 26917 & 6723 & 6412 & 125 & 1509
\enddata
\end{deluxetable*}

\begin{figure}
\includegraphics[width=0.48\textwidth,angle=0]{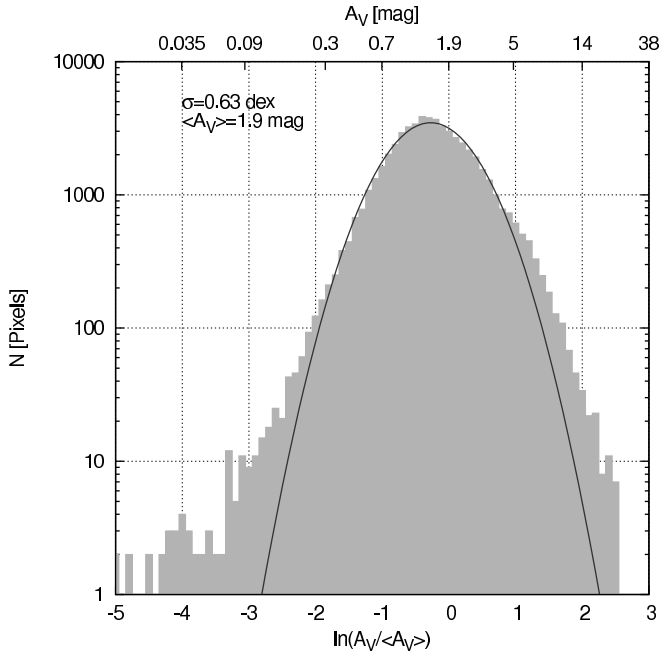}
     \caption{ Probability density function of the visual extinction
  in the Taurus molecular cloud. The solid line corresponds to a
  Gaussian fit to the distribution of the natural logarithms of
  $A_{\rm V}/\langle A_{\rm V}\rangle$. This fit considers only visual
  extinctions that are lower than 4.4\,mag, as the distribution
  deviates clearly from a Gaussian for larger visual extinctions.
  (see text). }\label{fig:lognormal}
\end{figure}

Numerical simulations have shown that the probability density function
(PDF) of volume densities in molecular clouds can be fitted by a
log-normal distribution. This distribution is found in simulations
with or without magnetic fields when self-gravity is not important
\citep{Ostriker2001,Nordlund1999,Li2004,Klessen2000}. A log-normal
distribution arises as the gas is subject to a succession of
independent compressions or rarefactions that produce multiplicative
variations of the volume density
\citep{Passot1998,Vazquez-Semadeni2001}. This effect is therefore
additive for the logarithm of the volume density. A log-normal
function can also describe the distribution of column densities in a
molecular cloud if compressions or rarefactions along the line of
sight are independent \citep{Ostriker2001,Vazquez-Semadeni2001}.  Note
that log--normal distributions are not an exclusive result of
supersonic turbulence as they are also seen in simulations with the
presence of self--gravity and/or strong magnetic fields but without strong
turbulence \citep{Tassis2010}.

Deviations from a log-normal in the form of tails at high or low
densities are expected if the equation of state deviates from being
isothermal \citep{Passot1998,Scalo1998}. This, however, also occurs in
simulations with an isothermal equation of state due to the effects of
self-gravity \citep{Tassis2010}.

\begin{figure}
\includegraphics[width=0.48\textwidth,angle=0]{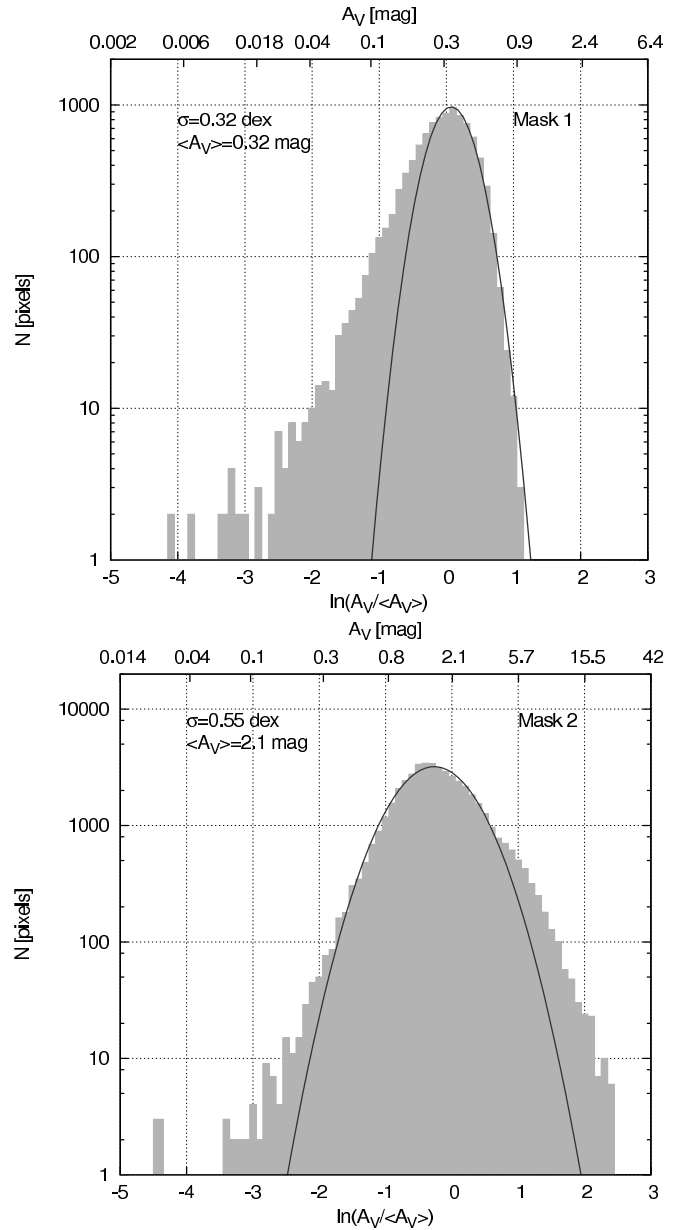}
     \caption{ Probability density function of the visual extinction
  for Mask\,1 ({\it upper panel}) and Mask\,2 ({\it lower panel}) in the Taurus molecular cloud. The solid
  line corresponds to a Gaussian fit to the distribution of the
  natural logarithm of $A_{\rm V}/\langle A_{\rm V}\rangle$. The fit for Mask1
  considers only visual extinctions that are larger than 0.24\,mag,
  while the fit for Mask\,2 includes only visual extinctions that are less
  than 4.4\,mag (see text). }\label{fig:lognormal_masks}
\end{figure}

\begin{figure}
\includegraphics[width=0.48\textwidth,angle=0]{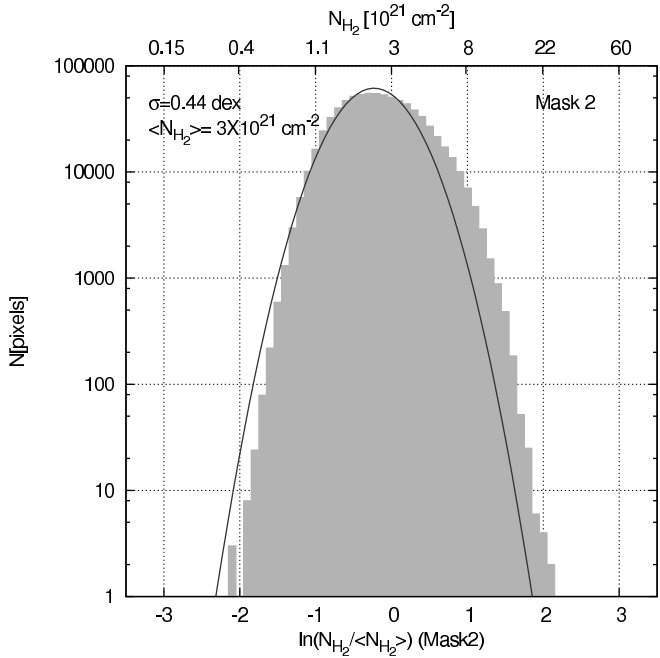}
     \caption{ Probability density function of the H$_2$ column
  density derived from $N({\rm CO})$ in Mask\,2 with an angular
  resolution of 47\arcsec\,(0.03\,pc at the distance of
  Taurus, 140\,pc). The solid line corresponds to a Gaussian fit to the
  distribution of the natural logarithm of $N({\rm H}_2)/\langle
  N({\rm H}_2)\rangle$. The fit considers H$_2$ column densities that
  are lower than 4$\times10^{21}$\,cm$^{-2}$ (or
  $\sim$4\,mag). }\label{fig:lognormal_nh2}
\end{figure}

In Figure~\ref{fig:lognormal} we show the histogram of the natural
logarithm of $A_{\rm V}$ in the Taurus molecular cloud normalized by
its mean value (1.9\,mag). Defining $x\equiv N/\langle N\rangle$,
where $N$ is the column density (either $A_{\rm V}$ or $N({\rm H}_2)$
), we fit a function of the form

\begin{equation}
f({\ln} x)=N_{\rm pixels} \exp \left [ -\frac{({\ln} (x) - \mu)^{2}}{2\sigma^2} \right ],
\end{equation}
where $\mu$ and $\sigma^2$ are the mean and variance of $\ln(x)$. The
mean of the logarithm of the normalized column density is related to the
dispersion $\sigma$ by $\mu=-\sigma^2/2$. In all Gaussian fits, we consider
$\sqrt N$ counting errors in each bin.

The distribution of column densities derived from the visual
extinction shows tails at large and small $A_{\rm V}$. The
large--$A_{\rm V}$ tail starts to be noticeable at visual extinctions
larger than $\sim$4.4\,mag. The low--$A_{\rm V}$ tail starts to be
noticeable at visual extinctions smaller than $\sim$0.26\,mag, which
is similar to the uncertainty in the determination of visual
extinction (0.29\,mag), and therefore it is not possible to determine
whether it has a physical origin or it is an effect of noise. The
distribution is well fitted by a log--normal for $A_{\rm V}$ smaller
than 4.4\,mag.   We searched in our extinction map for isolated
regions with peak $A_{\rm V} \gtrsim 4.4$\,mag. We find 57 regions
that satisfy this requirement. For each region, we counted the number
of pixels that have $A_{\rm V} \gtrsim 4.4$\,mag and from that
calculated their area, $A$.  We then determined their size using $L =
2\sqrt{(A/\pi)}$. The average value for all such regions is 0.41\,pc.
This value is similar to the Jeans length, which for $T_{\rm
kin}$=10\,K and $n({\rm H}_2)=10^{3}$\,cm$^{-3}$ is about 0.4 pc. This
agreement suggests that the high--$A_{\rm V}$ tail might be a result
of self--gravity acting in dense regions. \citet{Kainulainen2009}
studied the column density distribution of 23 molecular cloud
complexes (including the Taurus molecular cloud) finding tails at both
large and small visual extinctions.

\citet{Kainulainen2009} found that high--$A_{\rm V}$ tails are only
present in active star--forming molecular clouds while quiescent
clouds are well fitted by a log--normal.  We test whether this result
applies to regions within Taurus in Figure~\ref{fig:lognormal_masks}
where we show the visual extinction PDF for the Mask\,1 and 2
regions. Mask\,1 includes lines--of--sights that are likely of lower
volume density than regions in Mask\,2, and in which there is little
star formation.  This is illustrated in Figure~\ref{fig:masks} where
we show the distribution of the Mask regions defined in our map
overlaid by the compilation of stellar members of Taurus by
\citet{Luhman2006}. Most of the embedded sources in Taurus are located
in Mask\,2.  Note that the normalization of $A_{\rm V}$ is different
in the two mask regions. The average value of $A_{\rm V}$ in Mask\,1
is 0.32\,mag and in Mask\,2 is 2.1 mag. In Mask\,1 we see a tail for
low--$A_{\rm V}$ starting at about 0.2\,mag.  Again, this visual
extinction is close to the uncertainty in the determination of $A_{\rm
V}$.  For larger visual extinctions the PDF appears to be well fitted
by a log--normal distribution. In case of the visual extinction PDF in
Mask\,2, we again see the tail at large $A_{\rm V}$ starting at about
4.4\,mag. For lower values of $A_{\rm V}$ the distribution is well
represented by a log--normal.

We can use our CO map of Taurus at its original resolution
(47\arcsec\,which corresponds to 0.03\,pc at the distance of Taurus,
140\,pc) to study the column density PDF at higher resolution than the
200\arcsec\,$A_{\rm V}$ map (Figure~\ref{fig:lognormal_nh2}). We
estimate $N({\rm H}_2)$ from our CO column density map in Mask\,2 by
applying a constant [CO]/[H$_2$] abundance ratio of
1.1$\times$10$^{-4}$ (Section~\ref{sec:large-n12rm-co}).  The average
H$_2$ column density in Mask\,2 is $3\times10^{21}$\,cm$^{-2}$.  We do
not consider Mask\,1 because of the large scatter found in the
[CO]/[H$_2$] abundance ratio (Section~\ref{sec:low-n12rm-co}). In the
figure we see that the distribution is not well fitted by a
log--normal. As for $A_{\rm V}$, the PDF also shows a tail for large
column densities that starts to be noticeable at about
$4\times10^{21}$\,cm$^{-2}$ (or $A_{\rm V} \simeq 4$\,mag).
Therefore, the high--column density excess seems to be independent of
the spatial scale at which column densities are sampled.   We
repeated the procedure described above to search for isolated cores in
our map with $N({\rm H}_2) > 4\times10^{21}$\,cm$^{-2}$ and obtained
an average size for cores of 0.5\,pc, which is consistent to that
obtained in our $A_{\rm V}$ map .  Note that at this resolution we
are not able to account for effects of temperature gradients and of CO
depletion along the line of sight, as this requires knowledge of
$A_{\rm V}$ at the same resolution.  We therefore underestimate the
number of pixels in the H$_2$ column density PDF for $N({\rm H}_2)
\gtrsim 1\times10^{22}$\,cm$^{-2}$ while we overestimate them for
$N({\rm H}_2) \lesssim 1\times10^{22}$\,cm$^{-2}$. But the number of
pixels ($\sim$7000) affected by those effects represent only 9\% of
the number of pixels ($\sim$81000) that are in excess relative to the
log--normal fit between $3\times10^{21}$ and
$1\times10^{22}$\,cm$^{-2}$, and therefore the presence of a tail at
large--$N({\rm H}_2)$ is not affected.   Note that this also
affected our ability to identify isolated regions in the $N({\rm
H}_2)$ map. We were able to identify only 40 cores compared with the
57 found in the $A_{\rm V}$ map.

In summary, we find that the distribution of column densities in
Taurus can be fitted by a log--normal distribution but shows tails at
low and high--column densities. The tail at low--column density may be
due to noise and thus needs to be confirmed with more sensitive
maps. We find that the tail at large column densities is only present
in the region where most of the star formation is taking place in
Taurus (Mask\,2) and is absent in more quiescent regions
(Mask\,1). The same trend has been found in a larger sample of clouds
by \citet{Kainulainen2009}. Here we suggest that the distinction
between star-forming and non star-forming regions can be found even
within a single molecular cloud complex. The presence of tails in the
PDF in Taurus appears to be independent of angular resolution and is
noticeable for length scales smaller than 0.41\,pc.

\section{Conclusions}
\label{sec:conclusions}

In this paper we have compared column densities derived from the large
scale $^{12}$CO and $^{13}$CO maps of the Taurus molecular cloud
presented by \citealt{Narayanan2008} (see also
\citealt{Goldsmith2008}) with a dust extinction map of the same
region. This work can be summarized as follows,

\begin{itemize}

\item We have improved the derivation of the CO column density
compared to that derived by \citet{Goldsmith2008} by using an updated
value of the spontaneous decay rate and using exact numerical rather
than approximate analytical calculation of the partition function. We
also have used  data that has been corrected for error beam
pick--up using the method presented by \citet{Bensch2001}.

\item We find that in the Taurus molecular cloud the column density
and visual extinction are linearly correlated for $A_{\rm V}$ up to
10\,mag in the region associated with the B213--L1495 filament. In the
rest of Taurus, this linear relation is flattened for $A_{\rm V}
\gtrsim 4$\,mag. A linear fit to data points for $A_{\rm V} < 10$\,mag
and $N({\rm CO}) > 10^{17}$\,cm$^{-2}$ results in an abundance of CO
relative to H$_2$ equal to 1.1$\times$10$^{-4}$.

\item For visual extinctions larger than $\sim$4\,mag the CO column density
is affected by saturation effects and freezeout of CO molecules onto
dust grains. We find that the former effect is enhanced due to the
presence of edge--to--center temperature gradients in molecular
clouds. We used the RATRAN radiative transfer code to derive a
correction for this effect.

\item We combined the column density of CO in ice form derived from
observations towards embedded and field stars in Taurus by
\citet{Whittet2007} with the saturation--corrected gas--phase $N({\rm
CO})$ to derive the total CO column density (gas--phase+ices). This
quantity is linearly correlated with $A_{\rm V}$ up to the maximum
extinction in our data $\sim$23\,mag.

\item We find that the gas--phase CO column density is reduced by up
to a factor of $\sim$2 in high--extinction regions due to depletion in
the Taurus molecular cloud.

\item We fit an analytical column density profile to 13 cores in
Taurus. The mean value and standard deviation of the central volume density are
($1.4\pm0.4$)$\times10^{4}$\,cm$^{-3}$. We use the derived volume
density profile and the amount of depletion observed in each core to
derive an upper limit to the  CO depletion age with a mean value and standard deviation of
$(4.2\pm2.4)\times10^{5}$\,years. We find little variation of this age
among the  different regions within  Taurus.

\item For visual extinctions lower than 3\,mag we find that $N({\rm
CO})$ is reduced by up to two orders of magnitude due to the competition
between CO formation and destruction processes. There is a
large scatter in the $A_{\rm V}$ --$N({\rm CO})$ relation that is
suggestive of different FUV radiation fields characterizing the gas along
different lines--of--sight.

\item The mass of the Taurus molecular cloud is about
1.5$\times10^{4}\,$M$_{\odot}$. Of this, $\sim$49\% is contained in
pixels where both $^{12}$CO and $^{13}$CO are detected (Mask\,2),
$\sim$28\% where $^{12}$CO is detected but $^{13}$CO is not (Mask\,1),
and $\sim$23\% where neither $^{12}$CO nor $^{13}$CO are detected
(Mask\,0). 

\item We find that the masses derived from CO and $A_{\rm V}$ are in
  good agreement. For Mask\,2 and Mask\,0 we used a [CO]/[H$_2$]
  relative abundance of 1.1$\times$10$^{-4}$ and 1.2$\times10^{-6}$,
  respectively. For Mask\,1, we used a variable [CO]/[H$_2$] relative
  abundance taken from a fit to the average relation between $A_{\rm
  V}$ and $N({\rm CO})$ in this region, with $-6.7 < \log([{\rm
  CO}]/[\rm H_2]) \leq -3.9$.

\item We also compared the mass derived from $A_{\rm V}$ with the
$^{12}$CO $J = 1 \to 0$ luminosity for the regions derived above. For
Mask\,1 and 2 these two quantities are related with a CO--to--H$_2$
conversion factor of about 2.1$\times10^{20}$cm$^{-2}$ (K km s$^{-1}$
)$^{-1}$. The derived CO--to--H$_2$ conversion factor is in agreement
with that found in our Galaxy using $\gamma$--ray observations. In
Mask\,0, however, we find a larger the conversion factor of
1.2$\times10^{21}$cm$^{-2}$ (K km s$^{-1}$ )$^{-1}$.

\item We studied the distribution of column densities in Taurus. We
find that the distribution resembles a log--normal but shows tails at
large and low column densities. The length scale at which the
high--column density tail starts to be noticeable is about 0.4\,pc,
which is similar to the Jeans length for a $T$=10\,K and $n_{{\rm
H}_2}=10^{3}$\,cm$^{-3}$ gas, suggesting that self--gravity is
responsible for its presence. The high--column density tail is only
present in regions associated with star formation, while the more quiescent
positions  in Taurus do not show this feature.  This tail is independent
of the resolution of the observations. \\

\end{itemize}

\acknowledgments 

We would like to thank Douglas Whittet for the idea to add the column
density of CO--ices to the gas--phase CO column densities, Jonathan
Foster for providing sample extinction maps that were used to test the
implementation of the NICER algorithm used here, John Black, Edwine
van Dishoeck, and Ruud Visser for helpful discussions about the
formation/destruction processes affecting CO at low column densities,
specially Ruud Visser for providing results of his recent
calculations, Kostas Tassis for discussions about the nature of column
density distributions in molecular clouds, and Marko Kr\v{c}o for sharing
his H\,{\sc i} map of Taurus.  J.L.P was supported by an appointment
to the NASA Postdoctoral Program at the Jet Propulsion Laboratory,
California Institute of Technology, administered by Oak Ridge
Associated Universities through a contract with NASA.  This research
was carried out at the Jet Propulsion Laboratory, California Institute
of Technology and was supported by a grant from the National Science
Foundation.  This research has made use of NASA's Astrophysics Data
System Abstract Service.

\appendix

\section{Error Beam Correction}
\label{sec:error-beam-corr}

The FCRAO 14m telescope is sensitive not only to emission that couples
to the main beam (with efficiency $\eta_{\rm mb}$~=~0.45 at 115~GHz
and $\eta_{\rm mb}$~=~0.48 at 110~GHz) but also to emission
distributed on scales comparable to the error beam (30\arcmin). For
emission extended over such large-scale, the coupling factor
(including the main beam contribution) is the forward spillover and
scattering efficiency, $\eta_{\rm fss}$~=~0.7, at both frequencies. The
error beam pickup, also known as ``stray radiation'', can complicate
the accurate calibration of the measured intensities: a
straightforward scaling of the data by 1/$\eta_{\rm mb}$ can
significantly overestimate the true intensity in regions where
emission is present on large angular scales. Given the wide range of
angular sizes of the structures in Taurus, it is clear in general that
neither $\eta_{\rm mb}$ nor $\eta_{\rm fss}$ will give optimum
results.


To accurately scale the FCRAO data, it is essential to remove the
error beam component before scaling the intensities to the main beam
scale.  Methods for correcting millimeter-wave data for error beam
pickup have been discussed by \citet{Bensch2001}, who introduce the
``corrected main beam temperature scale'' ($T_{\rm mb,c}$) with which
optimum calibration accuracy is achieved by scaling the data by
1/$\eta_{\rm mb}$ {\it after} removal of radiation detected by the
error beam.

To remove the error beam component we use the second of the methods described in
\citet{Bensch2001}. The error beam component is
removed in Fourier space directly from the FCRAO data, and the intensities are
converted to the $T_{\rm mb,c}$ scale, by the following method:

\begin{itemize}

\item The Fourier transform of the antenna temperature is taken: $\tilde{T}_{\rm A}^{*}$~=~$FT(T_{\rm A}^{*})$

\item The following correction is applied to each velocity slice in the cube: 
$$\tilde{T}_{\rm mb,c} = \tilde{T}_{\rm A}^{*}
(\eta_{\rm mb}+\eta_{\rm eb}\exp(\frac{-\pi^{2}(\theta_{\rm eb}^{2}-\theta_{\rm mb}^{2})(k_{x}^{2}+k_{y}^{2})}{4\ln(2)}))^{-1}$$
where $\theta_{\rm eb}$ is the FWHM of the error beam, $\theta_{\rm mb}$ is
the FWHM of the main beam, $\eta_{\rm  eb} = \eta_{\rm  fss} - \eta_{\rm mb}$, and
$k_{x}$, $k_{y}$ are the wavenumbers along the $x$, $y$ (RA, decl.)
directions respectively.

\item The inverse Fourier transform is performed, with only the real
part of the result being retained: $T_{\rm mb,c}$~=~$Re~(IFT(\tilde{T}
_{\rm mb,c}))$.   The imaginary part is consistent with round--off
errors.

\end{itemize}

\begin{figure}
\includegraphics[width=\textwidth,angle=0]{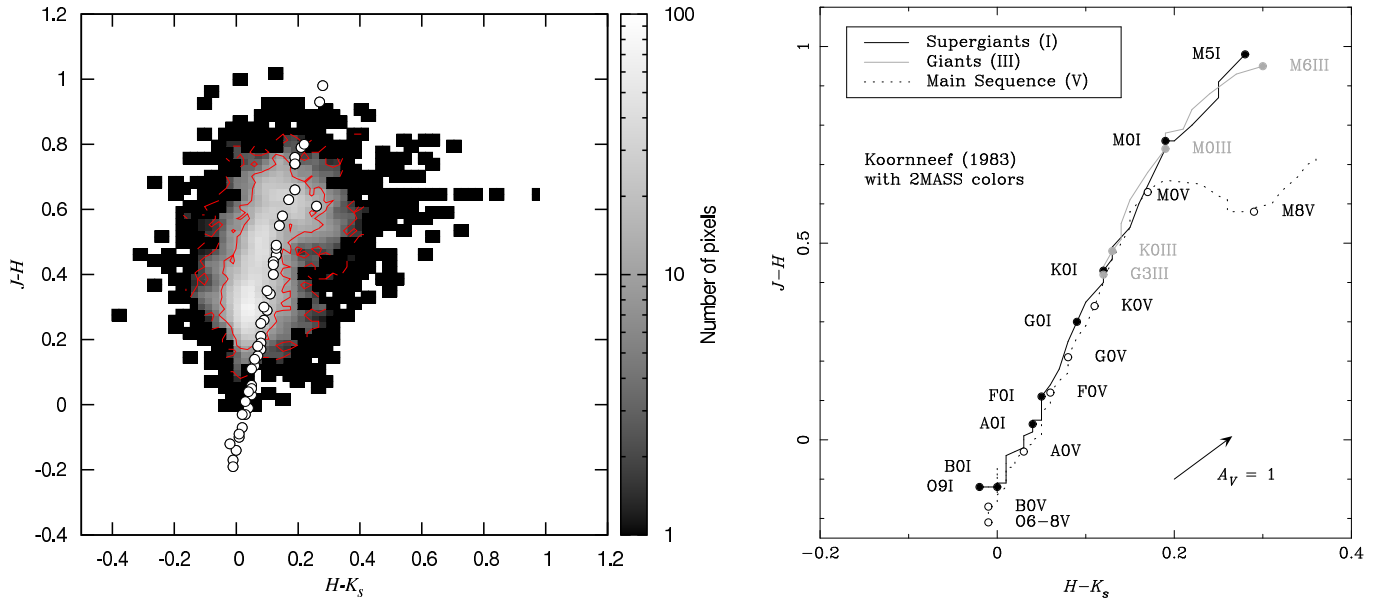}
     \caption{ { \it (left)} Color $(H-K_s)$ versus $(J-H)$ for stars
     observed in the control field.  {\it (right}) Intrinsic $J-H$ and
     $H-K_s$ colors of Main Sequence, Giant, and Supergiant stars
     (taken from \citealt{Koorneef1983}). These stars are indicated by
     circles in the left panel.  }\label{fig:2}
\end{figure}

At low spatial frequencies, the correction factor is
$\sim$~1/$\eta_{\rm fss}$ while at high spatial frequencies, the
correction factor is $\sim$~1/$\eta_{\rm mb}$. The effective
correction factor at any point is determined therefore by the spatial
structure of the emission in the vicinity of that point.  More
detailed information and quantitative analysis of the above procedure
can be found in Brunt et al (2010, in prep) and Mottram \& Brunt
(2010, in prep). For typical applications, a naive scaling by
1/$\eta_{\rm mb}$ overestimates the true intensities, as inferred from
comparison to CfA survey data \citep{Dame01}, by around 25--30\%. For
reference, an overestimation of $\sim$~50\% would be applicable if
$\eta_{\rm fss}$ were appropriate everywhere. The spatially variable
correction factor afforded by the method used here therefore offers a
higher fidelity calibration of the data.

\section{The Extinction Map}
\label{sec:extinction-map}

We have used the 2MASS point source catalog to create an
near--infrared extinction map of Taurus. This was done using an
implementation of NICER \citep{Lombardi2001} from \citet{Chapman2007}.
The 2MASS catalog we used has 1039735 ($\sim$1 million) stars over an
area between RA=04:03:51.6 and 05:05:56.6 and decl.=+19:24:14.4 and
+30:50:24 (J2000). We use the compilation by \citet{Luhman2006} to
remove 156 stars that are known to be  members of Taurus. The
map generated has an angular resolution of 200\arcsec\,and is Nyquist
sampled with a pixel spacing of 100\arcsec, corresponding to 0.07\,pc
at a distance of 140\,pc. The resolution of the map was determined by
that of the H\,{\sc i} map used to correct the data for the
contribution of H\,{\sc i} to the total extinction (see below). The
final extinction map is shown in Figure~\ref{fig:maps_all}.

We constructed extinction maps in nearby regions around Taurus with
the goal of finding a field that does not show significant extinction,
so it can be used as a control field to estimate the intrinsic $(J-H)$
and $(H-K_s)$ stellar colors. We selected a region  corresponding to
a $2^{\circ}\times2^{\circ}$ box centered at RA=03:50:44.7 and
decl.=+27:46:54.1 (J2000).  The mean ($\pm$ weighted standard
deviation) values for stars in this box are 0.454$\pm$0.157 mag for
$(J-H)$ and 0.114$\pm$0.074 mag for $(H-K_s)$.  We also computed the
covariance matrix for the $(J-H)$ and $(H-K_s)$ colors.  The on-axis
elements of this matrix are $\sigma_{J-H}^2$ and $\sigma_{H-K_s}^2$,
the dispersions of the $(J-H)$ and $(H-K_s)$ colors in the control field,
while the two off--axis elements are identical to each other, with a
value of 0.006.  In Figure~\ref{fig:2} we show the color $(H-K_s)$
versus $(J-H)$ of stars in the control field. We also show the
intrinsic color of Main Sequence, Giant, and Supergiant stars. Apart
from the scatter due to photometric errors, there is large scatter in
the intrinsic colors due to different stellar types in the control
field.  The mean values, weighted standard deviation and off--axis
covariance  matrix are input to the NICER routine and with them we
correct for the different sources of scatter of the intrinsic colors
in the control field.  Note that \citet{Padoan2002} used an intrinsic
$(H-K_s)$ color of 0.13 mag in their extinction map of Taurus. The
difference relative to that in  our control field is thus seen to be small.

 We transformed the $(J-H)$ and $(H-K_s)$ colors to $A_{\rm V}$ using
an extinction curve from \citet{Weingartner2001} with a ratio of
selective to total extinction $R_{\rm V}$=3.1.  Note that this $R_{\rm
V}$ is derived towards diffuse regions ($A_{\rm V} \leq $ 1.5 mag). At
larger volume densities the value of $R_V$ is expected to increase up
to 4.5 at the center of dense cores due to grain growth by accretion
and coagulation \citep{Whittet2001}. Considering that a given
line--of--sight might intersect both dense and diffuse regions,
\citet{Whittet2001} estimated an effective $R_{\rm V}$ that increases
up to $\sim 4.0$ for $A_{\rm V}$ $\simeq 10$\,mag. For such a value of
$A_{\rm V}$ we expect that for a given total hydrogen column density,
the visual extinction increases by about 20\% due to enhanced
scattering as the grain sizes increases \citep{Draine2003}.

\begin{figure}
  \includegraphics[width=\textwidth,angle=0]{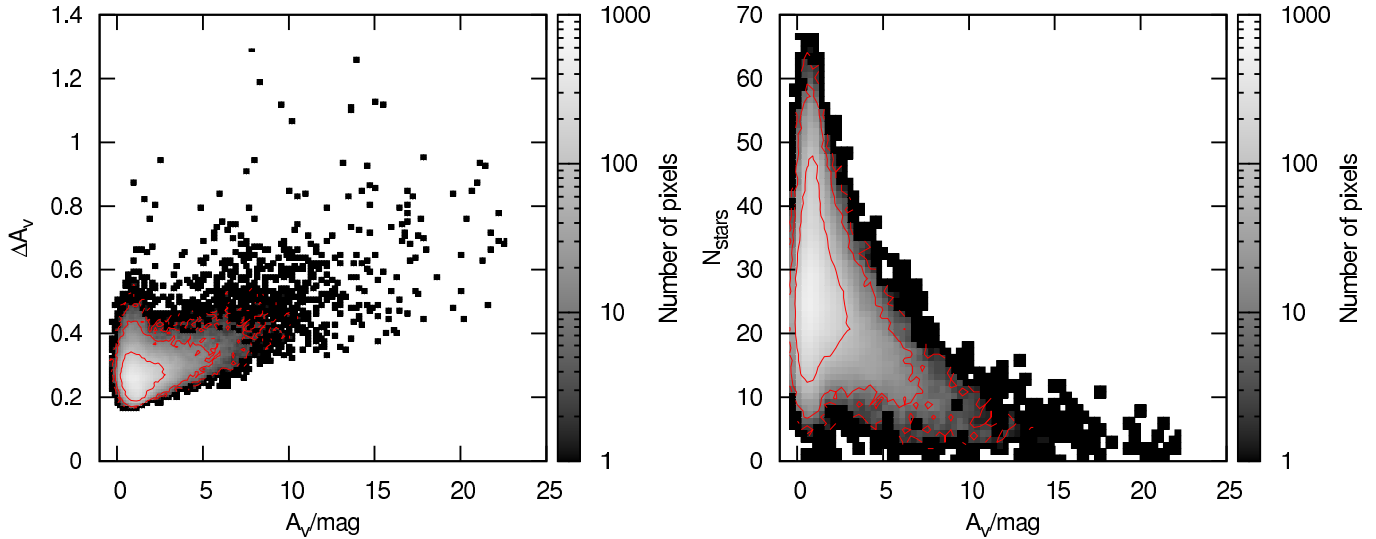}
     \caption{The formal error per pixel ({\it left}) and the number
of stars per pixel ({\it right}) as a function of the visual
extinction estimated in the Taurus molecular
cloud. }\label{fig:av_parameters}
\end{figure}

Figure~\ref{fig:av_parameters} shows the relation between the
estimated visual extinction in the Taurus molecular cloud and the
formal error per pixel (i.e. error propagation from the error in the
estimation of $A_{\rm V}$ for each star) and the number of stars per
pixel.  The errors in $A_{\rm V}$ range from $\sim$0.2\,mag at low
extinctions to $\sim$1.3\,mag at large visual extinctions. The average
error is 0.29\,mag while the average number of stars per pixel is 28.
As expected, the number of stars per pixel decreases as the extinction
increases.

\begin{figure*}
\centering
  \includegraphics[width=0.45\textwidth,angle=0]{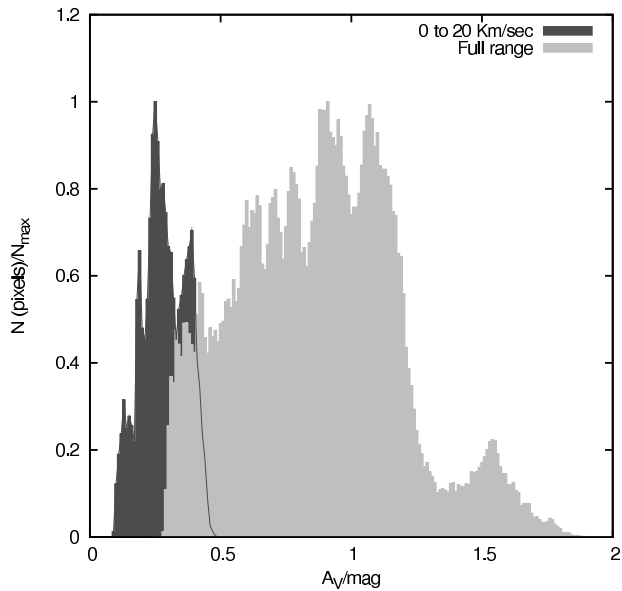}
     \caption{Histogram of the visual extinction in Taurus associated
     with H\,{\sc i} for the full range of 21\,cm velocities ({\it
     light gray}) and for the velocity range between 0 and 20 km
     s$^{-1}$ ({\it dark gray}). The histogram has been determined for
     the region where both $^{12}$CO and $^{13}$CO are
     detected. }\label{fig:mask2_hist}
\end{figure*}

There is a large filamentary H{\sc i} structure, extending away from
the Galactic plane, which coincides with the eastern part of Taurus.
Based on distances of molecular clouds at the end of the filament, we
assume that this filament lies between the Taurus background stars and
the Earth. Dust in the filament will thus contribute to the total
extinction measured.  We estimate the contribution to the visual
extinction from dust associated with H\,{\sc i} using the Arecibo map
from Marco Kr\v{c}o (PhD Thesis, Cornell University, in preparation). In Figure~\ref{fig:mask2_hist} we show a
histogram of the visual extinction associated with positions in the
H\,{\sc i} map where both $^{12}$CO and $^{13}$CO are detected in our
Taurus map.  We show the extinction for the full range of velocities
and for the range between 0 to 20 km s$^{-1}$ (similar to the velocity
range at which CO emission is observed).  We correct the $A_{\rm V}$
map by extinction associated with neutral hydrogen in the latter range
(see below). The average correction is $\sim$0.3\,mag.

\begin{figure}
  \includegraphics[width=\textwidth,angle=0]{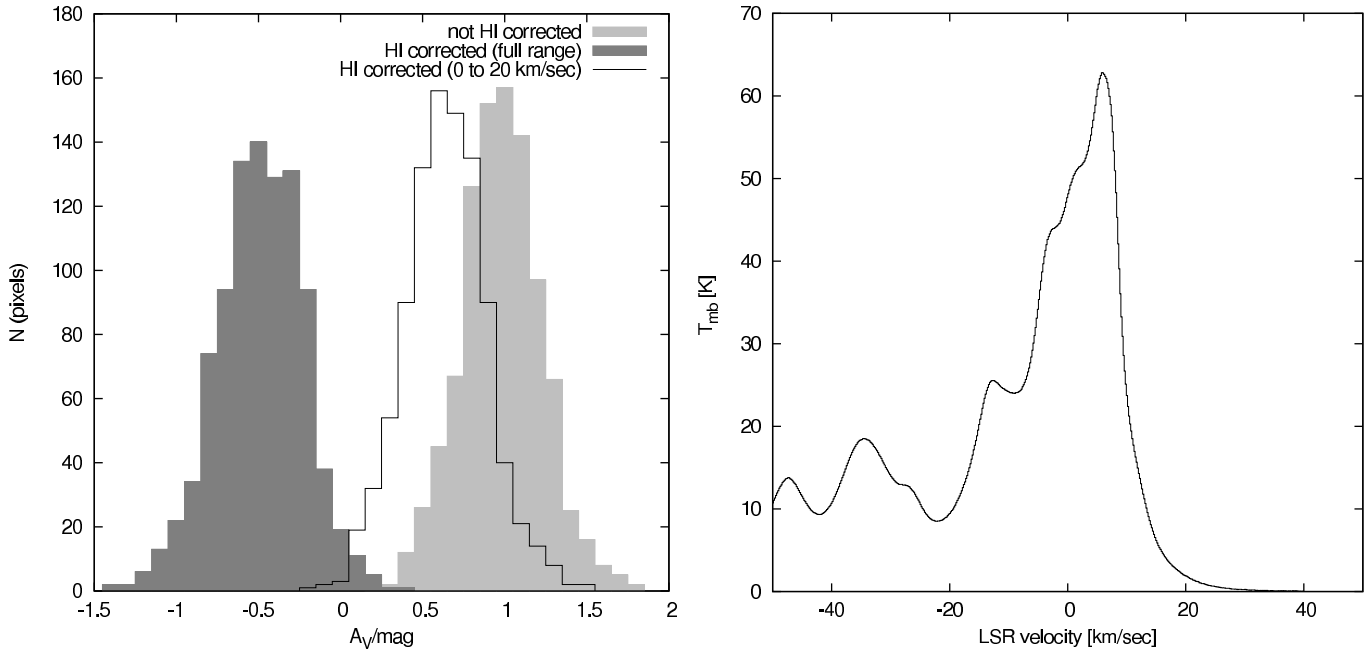}
     \caption{({\it left}) Histogram of the visual extinction calculated
in a field with complex velocity structure (see text) with and without correction for extinction
associated with H\,{\sc i}.  ({\it right}) The H\,{\sc i} spectrum averaged
over the this field. } \label{fig:field_hi}
\end{figure}

In order to see whether some H\,{\sc i} velocity components are
foreground to the 2MASS stars, we examine a field with complex H\,{\sc
i} velocity structure. We choose a region northwest of the Taurus
molecular cloud that shows small visual extinction (RA = 04:57:25.472
and decl. = 29:07:0.81). Figure~\ref{fig:field_hi}a shows a histogram
of the visual extinction without correction, corrected for H\,{\sc i}
over the entire velocity range, and corrected for H\,{\sc i} over the
0 to 20 km s$^{-1}$ range.  We also show in Figure~\ref{fig:field_hi}b
the average H\,{\sc i} spectrum in the selected field. The negative
velocity components produce significant excess reddening associated
with H\,{\sc i} that is inconsistent with the extinction determined
from 2MASS stars. We therefore conclude that H\,{\sc i} components
with negative velocities are background to the 2MASS stars.  This
confirms the correctness of excluding negative velocities for
determining the H\,{\sc i}-associated extinction correction for
Taurus. The exact velocity range used is a source of uncertainty of a
few tenths of a magnitude in the extinction.

We finally note that the widespread H\,{\sc i} emission is also
present in the control field.  The control field is contaminated by
$\sim$0.12\,mag of visual extinction associated with H\,{\sc i}.  This
contribution produces a small overestimation of the intrinsic colors
in the control field. Therefore, since we determine visual extinctions
based on the difference between the observed stellar colors in Taurus
and those averaged over the control field, we have added  0.12\,mag to our
final $A_{\rm V}$ map of Taurus.

\section{Correction for Temperature Gradients along the Line--of--Sight}
\label{sec:radi-transf-model}

In order to assess the impact of core-to-edge temperature gradients in
the estimation of $N({\rm CO})$, we use the radiative transfer code
{\tt RATRAN} \citep{Hogerheijde2000}. With {\tt RATRAN} we calculate
$^{12}$CO and $^{13}$CO line profiles and integrated intensities from
a model cloud and use them to estimate $N({\rm CO})$, following our
analytic procedure (Section~\ref{sec:nh_2-map-derived}),  which we then
compare with that of the original model cloud.

\begin{figure*}[h]
\centering
\includegraphics[width=\textwidth,angle=0]{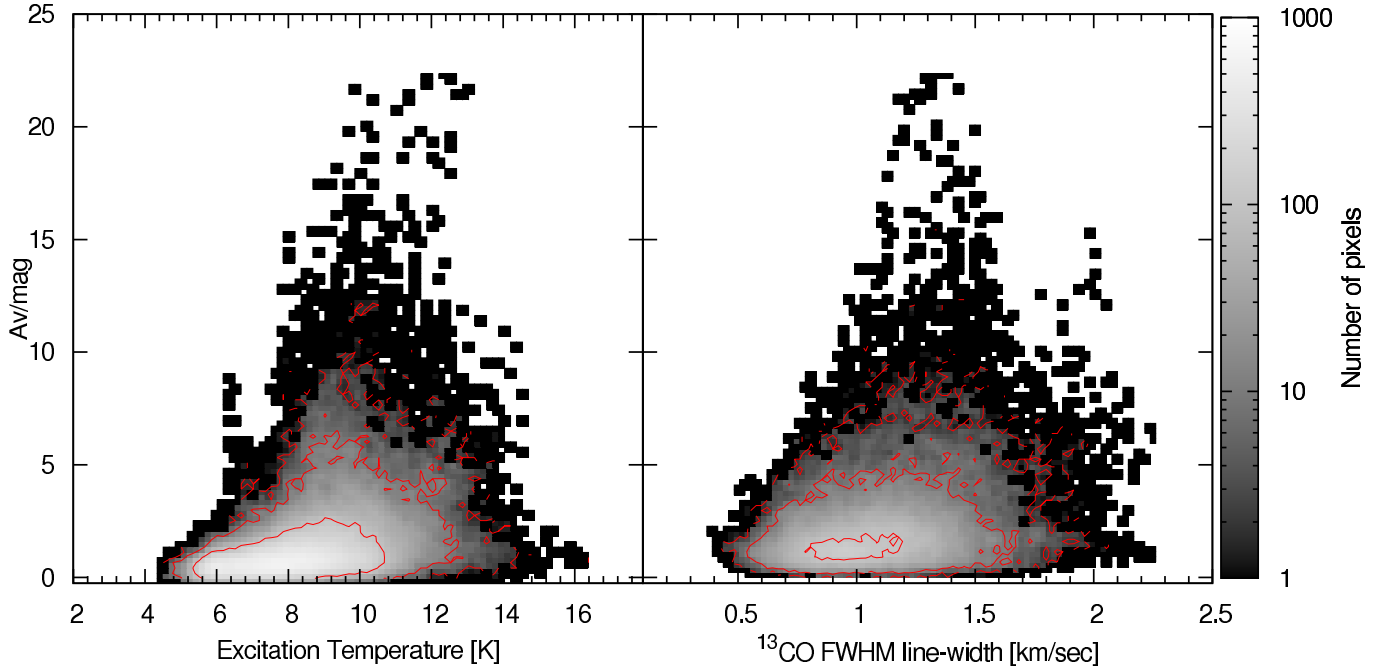}
\caption{Pixel-by-pixel comparison between the visual extinction
($A_{\rm V}$) and the $^{12}$CO-derived excitation temperature ({\it
left }) and $^{13}$CO FWHM line--width ({\it right}).
}\label{fig:tex_av}
\end{figure*}

The model is a spherical cloud with a truncated power-law
density profile.  We adopt a density profile
$n(r)$=$n_\mathrm{s}(r/r_\mathrm{c})^{-\alpha}$ for 0.3$r_\mathrm{c}$
$\leq$ $ r \leq r_\mathrm{c}$, and constant density,
$n(r)=n_s$(0.3)$^{-\alpha}$ in the central portion of the cloud ($r <
0.3 r_\mathrm{c}$).  Here, $r_\mathrm{c}$ is the cloud radius and
$n_\mathrm{s}$ the density at the cloud surface.  In our models we use
a power--law exponent of $\alpha$ = 1.96 for the density profile, a
density at the cloud surface of 2$\times$10$^{4}$~cm$^{-3}$, and a
cloud radius of 5$\times$10$^{16}$~cm (0.015\,pc at a distance of
140\,pc). The values of the density at the cloud surface and the
power--law exponent are taken from the fit to the extinction profile
of the prestellar core B68 \citep{Alves01}.  We adopt a FWHM
line-width in the model of $\Delta v$=1.33~km~s$^{-1}$, in order to
have line widths that are consistent with those in pixels having
$A_{\rm V} > $ 10 mag (Figure~\ref{fig:tex_av}).  We use the $^{13}$CO
and $^{12}$CO emission resulting from the radiative transfer
calculations to derive, using Equation (\ref{eq:9}), the CO column
density ($N({\rm CO})_{\rm emission}$) that will be compared with that
of the model cloud ($N({\rm CO})_{\rm model}$).  We trace CO column
densities between 3$\times10^{16}$\,cm$^{-2}$ and
1$\times10^{18}$\,cm$^{-2}$.

We consider the case of isothermal clouds and of clouds with
temperature gradients. Temperature gradients can be produced, for
example, when clouds are externally illuminated by the interstellar
radiation field \citep[e.g.][]{Evans2001}. We adopt a temperature
profile given by

\begin{equation}
T(r)=(T_{\rm s}-T_{\rm c})\left (\frac{r}{r_{\rm c}} \right)^2+T_{\rm c},
\end{equation}
where $T_{\rm c}$ and $T_{\rm s}$ are the temperature at the cloud
center and surface, respectively.

We run isothermal cloud models with kinetic temperatures of 8, 9,
10,12, and 15\,K. In the case of clouds with temperature gradients we
consider the same range of temperatures for the cloud surface and
$T_{\rm c}=8$\,K for the cloud center. We choose this value because
the balance between the dominant heating and cooling mechanisms in
dense and shielded regions, namely cosmic-ray heating and cooling by
gas-grain collisions, typically results in this range of temperatures
\citep{Goldsmith2001}. The selected range of temperatures match the
observed range of excitation temperatures for $A_{\rm V} > 10$\,mag
(Figure~\ref{fig:tex_av}).


\begin{figure*}
\centering
\includegraphics[width=0.85\textwidth,angle=0]{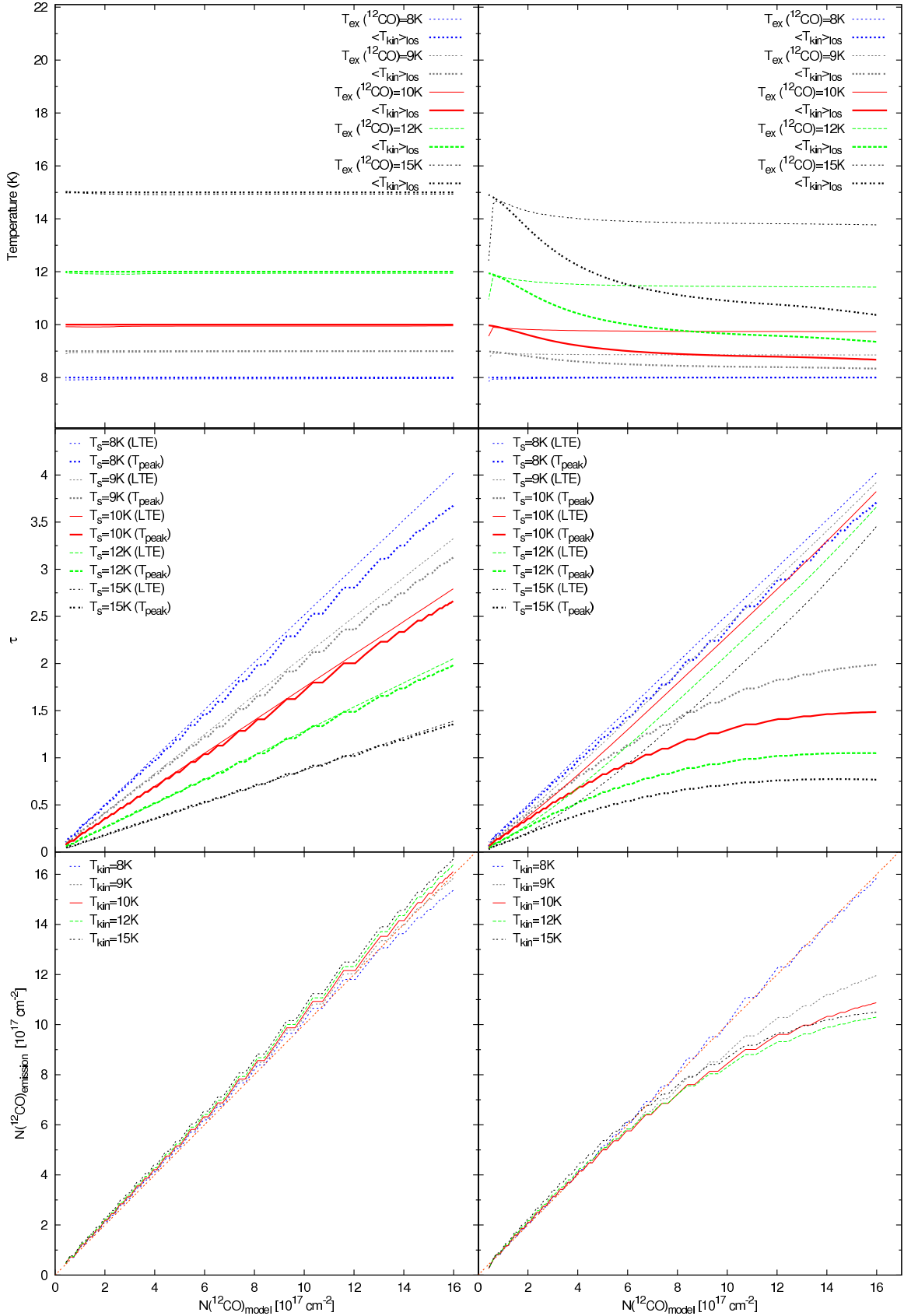}
\caption{ ({\it upper row}) Excitation temperature derived from the
     $^{12}$CO emission as function of $N({\rm CO})_{\rm model}$ for
     isothermal cloud models ({\it left }) and models with temperature
     gradients ({\it right}). In both panels we also show the model
     cloud kinetic temperature averaged along the line of sight,
     $\langle T_{\rm kin}\rangle_{\rm los}$, as a function of $N({ \rm CO})_{\rm
     model}$.  ({\it middle row}) The $^{13}$CO opacity versus $N({
     \rm CO})_{\rm model}$ for the models shown in the upper
     row. The opacity was derived using Equation~(\ref{eq:4}). We
     also show the opacity calculated from the model cloud
     N($^{13}$CO) with the assumption of LTE. ({\it lower row})
     Column density of CO calculated from the $^{12}$CO and $^{13}$CO
     emission using Equation~(\ref{eq:9}) versus the model cloud
     $N({\rm CO})$. The straight line corresponds to an one--to--one
     relation.}\label{fig:temp_grad_isrf}
\end{figure*}  

In the upper rows of Figure~\ref{fig:temp_grad_isrf} we show the
excitation temperature derived from the model $^{12}$CO emission as a
function of $N({\rm CO})_{\rm model}$. The left panel corresponds to
isothermal clouds and the right panel to clouds with temperature
gradients. We also show the line-of-sight (LOS) averaged kinetic
temperature as a function of $N({\rm CO})_{\rm model}$. In clouds
with temperature gradients, low column densities are on average warmer
than larger column densities. For isothermal clouds $T_{\rm ex}$ and
the model kinetic temperature are almost identical for all values of
$N({\rm CO})_{\rm model}$.  In the case of clouds with temperature
gradients we see that, although the average LOS kinetic temperature
decreases for large CO column densities, the derived excitation
temperature shows little variation with $N({\rm CO})_{\rm model}$,
tracing only the temperature at the cloud surface. This is a result of
$^{12}$CO becoming optically thick close to the cloud surface and
therefore the  $T_{\rm ex}$ determined in this manner applies only to this
region. In clouds with temperature gradients, using $^{12}$CO to
calculate the excitation temperature overestimates its value for
regions with larger column densities, where most of the $^{13}$CO
emission is produced.

We show the $^{13}$CO opacity (Equation~[\ref{eq:4}]) for both
isothermal clouds and clouds with temperature gradients as a function
of $N({\rm CO})_{\rm model}$ in the middle panels of
Figure~\ref{fig:temp_grad_isrf}.  We also show opacities calculated
from the model cloud $^{13}$CO column densities assuming LTE
($\tau_{\rm LTE}$).  When the cloud kinetic temperature is constant,
both opacities show good agreement for all sampled values of $N({\rm
CO})_{\rm model}$. In contrast, for clouds with temperature
gradients, opacities derived from the model line emission are lower
than $\tau_{\rm LTE}$ by up to a factor of $\sim$3. The differences
arise due to the overestimation of the excitation temperature in
regions with large $N({\rm CO})$, as Equation~[\ref{eq:4}] assumes a
constant value of $T_{\rm ex}$.

The relation between $N({\rm CO})_{\rm emission}$ and $N({\rm
CO})_{\rm model}$ is shown in the lower panels of
Figure~\ref{fig:temp_grad_isrf}. Isothermal clouds show almost a
one-to-one relation between these two quantities whereas clouds with
temperature gradients show that the relation deviates from linear for
large $N({\rm CO})_{\rm model}$.  This is produced by the
underestimation of opacities that affect the correction for this
quantity (Equation~[\ref{cf1}]).  The difference between $N({\rm
CO})_{\rm emission}$ and the expected CO column density is about 20\%.

\begin{figure*}
\centering
\includegraphics[width=0.5\textwidth,angle=0]{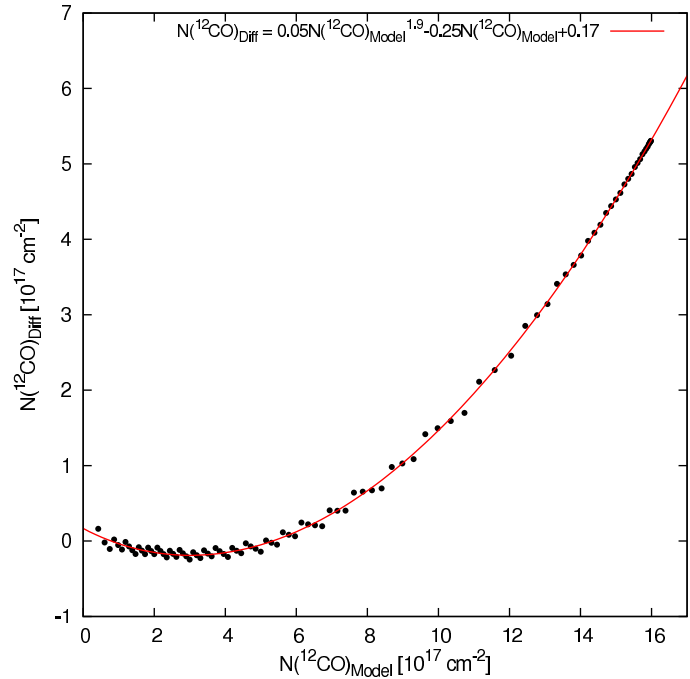}
\caption{Difference between the expected $N({\rm CO})_{\rm model}$ and
the derived $N({\rm CO})_{\rm emission}$ as a function of the model CO
column density. The red line represents a second-order polynomial
fit. }\label{fig:correction_function}
\end{figure*}

In the following we use the relation between $N({\rm CO})_{\rm model}$
and $N({\rm CO})_{\rm emission}$ to derive a correction to the
observed CO column densities. We notice that the difference between
these quantities does not show a strong dependence in the cloud
surface temperature. This is because all models have the same
temperature at the cloud center. Since the observed excitation
temperatures lie between $\sim9-15$\,K for $A_{\rm V} > 10$\,mag
(Figure~\ref{fig:temp_grad_isrf}), and the excitation temperature
derived from $^{12}$CO is similar to the kinetic temperature at the
cloud surface, we average all models from $T_{\rm s}=9\,$K to 15\,K in
steps of 1\,K to derive a correction function. In
Figure~\ref{fig:correction_function} we show the difference between the
model and derived CO column density ($N({\rm CO})_{\rm diff} = N({\rm
CO})_{\rm model}-N({\rm CO})_{\rm emission}$) as a function of the
model CO column density. To this relation we fit a polynomial function
given by

\begin{equation}\label{eq:11}
\frac{N({\rm CO})_{\rm diff}}{10^{17}{\rm cm}^{-2}}=0.05 N({\rm CO})_{\rm model}^{1.9}-0.25 N({\rm CO})_{\rm model}+0.17
\end{equation}

To apply this correction we made a rough estimate of the gas--phase
CO column density as a function of visual extinction. We use the
observations by \citet{FLW1982} of C$^{18}$O and the rarer isotopic
species C$^{17}$O and $^{13}$C$^{18}$O in the direction of field stars
located behind Taurus. We convert the observed column densities into
$N({\rm CO})$ assuming [CO]/[C$^{18}$O]=557,
[C$^{18}$O]/[C$^{17}$O]=3.6, and [C$^{18}$O]/[$^{13}$C$^{18}$O]=69
\citep{Wilson1999}.  Due to their low abundances, these species are
likely not affected by saturation.   Note that, however, they are
still sensitive to the determination of the excitation
temperature. \citet{FLW1982} presented column densities as lower
limits when $^{12}$CO is used to determine $T_{\rm ex}$($^{12}$CO)
(average $\sim$10\,K) and as upper limits when they used $T_{\rm
ex}$($^{12}$CO)/2 (i.e. $\sim$5\,K) as the excitation
temperature. The kinetic temperature in dense regions is likely to be
in between 5 and 10\,K \citep{Goldsmith2001} and therefore, assuming
that the isotopologues are thermalized, the excitation temperature
should also have a value in this range. Thus, we use the
average value between upper and lower limits of the CO column density
to determine its relation with $A_{\rm V}$. For the visual extinction
at the positions observed by \citet{FLW1982}, we use updated values
derived by \citet{Shenoy2008} from infrared observations\footnote{Note
that they used a relation between visual extinction and infrared color
excess determined in Taurus of $A_{\rm V}/E_{J-K}\simeq 5.3$
\citep{Whittet2001} which differs from that determined in the diffuse
ISM ($A_{\rm V}/E_{J-K}\simeq 6$). }.  We note that the visual
extinction correspond to a single star while the \citet{FLW1982}
observations are averaged over a 96\arcsec\,beam.  We constructed an
extinction map of Taurus with 96\arcsec\,resolution and an extinction
curve that matches that adopted by \citet{Shenoy2008} in order to
compare with their determination of $A_{\rm V}$. We found that the
visual extinctions always agree within $\pm0.4$\,mag.

\begin{figure*}
\includegraphics[width=1\textwidth,angle=0]{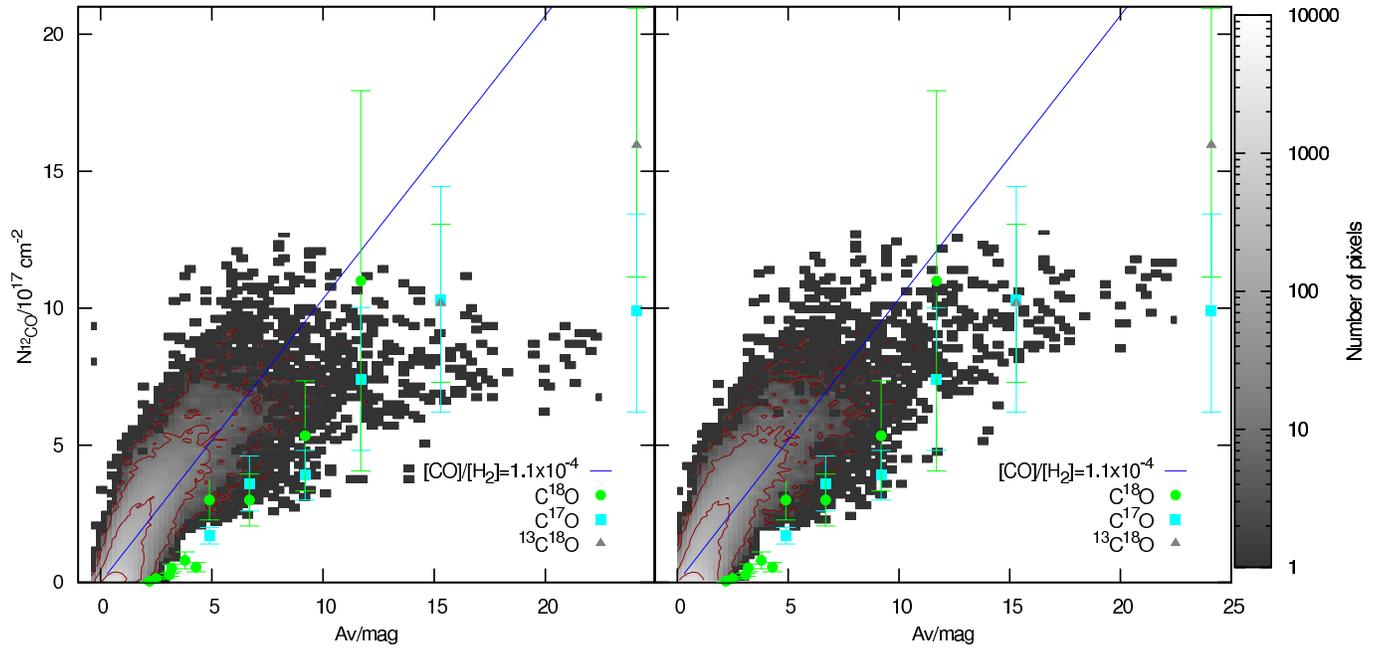}
\caption{ Pixel-by-pixel comparison between $A_{\rm V}$  and
$N({\rm CO})$ as in Figure~\ref{fig:av_nh2_all} including the CO
column density derived from rare isotopic species observed by
\citet{FLW1982}. The left panel shows the CO column densities derived
from Equation (\ref{eq:9}) with the opacity correction from
Equation~(\ref{cf1}), which assumes isothermal gas. The right panel
shows $N({\rm CO})$ corrected for saturation including temperature gradients.  }\label{fig:nco_av_cf_corr}
\end{figure*}

In Figure~\ref{fig:nco_av_cf_corr} we show the relation between
$N({\rm CO})$ and $A_{\rm V}$ with and without the correction for the
effects of temperature gradients along the line--of--sight. For
reference we include the values of $N({\rm CO})$ derived from the
observations by \citet{FLW1982}. The error bars denote the upper and
lower limits to the CO column density mentioned above.  Although our
determination of the gas--phase $N({\rm CO})$/$A_{\rm V}$ relation is
necessarily approximate, the validity of the correction for the effects
of temperature gradients along the line--of--sight is confirmed by the
good agreement between $A_{\rm V}$ and $N({\rm CO})$ up to $A_{\rm
V}\simeq$23\,mag after the addition of the column density of CO--ices
(Section~\ref{sec:co-depletion-1}).

\bibliography{ms}
\bibliographystyle{apj.bst}


\clearpage

\end{document}